\documentclass[12pt]{article}
\usepackage{epsfig}
\usepackage{axodraw}

\newcommand{\xF}{x_{\mathrm{F}}}
\newcommand{\pt}{p_{\perp}}
\newcommand{\kt}{k_{\perp}}
\renewcommand{\c}{\mathrm{c}}
\renewcommand{\d}{\mathrm{d}}
\newcommand{\e}{\rm{e}}
\newcommand{\ee}{\rm{e}^+\rm{e}^-}
\newcommand{\g}{\mathrm{g}}

\newcommand{\p}{\mathrm{p}}
\newcommand{\q}{\mathrm{q}}
\renewcommand{\u}{\mathrm{u}}
\renewcommand{\d}{\mathrm{d}}
\newcommand{\s}{\mathrm{s}}
\newcommand{\D}{\mathrm{D}}

\newcommand{\cbar}{\overline{\mathrm{c}}}

\newcommand{\qbar}{\overline{\mathrm{q}}}
\newcommand{\ubar}{\overline{\mathrm{u}}}

\newcommand{\Dbar}{\overline{\mathrm{D}}}

\newcommand{\Py}{{\sc{Pythia}}}

\newenvironment{Itemize}{\begin{list}{$\bullet$}%
{\setlength{\topsep}{0.2mm}\setlength{\partopsep}{0.2mm}%
\setlength{\itemsep}{0.2mm}\setlength{\parsep}{0.2mm}}}%
{\end{list}}
\newcounter{enumct}

\newlength{\abstwidth}
\begin{document}

\sloppy

\pagestyle{empty}

\begin{flushright}
LU TP 98--24\\
November 1998
\end{flushright}
 
\vspace{\fill}

\begin{center}
{\LARGE\bf Heavy Quark Fragmentation}\\[10mm]
{\Large E. Norrbin\footnote{emanuel@thep.lu.se}} \\[3mm]
{\it Department of Theoretical Physics,}\\[1mm]
{\it Lund University, Lund, Sweden}
\end{center}
 
\vspace{\fill}
 
\begin{center}
{\bf Abstract}\\[2ex]
\begin{minipage}{\abstwidth}
I present the main aspects of open charm production in $\pi^-\p$ collisions in the
context of the Lund String Fragmentation Model as implemented in the
Monte Carlo program \Py. The emphasis is on the transition from large to small
strings and the dependence on model parameters. A modified version is presented
and compared with experimental results both on asymmetries, single-charm spectra
and correlations.

\end{minipage}
\end{center}
 
\vspace{\fill}

\clearpage
\pagestyle{plain}
\setcounter{page}{1}

\section*{Introduction}

Since the discovery of charm in 1974 numerous experiments on charm production
at fixed target have been performed \cite{oldobs}.
In recent years the precision has increased significantly \cite{WA82,E769,E791},
which enables a detailed comparison with theory. Perturbative QCD can
describe some of the phenomenology of charm production, but not all
\cite{pertcharm1,pertcharm2}.
Most notably the asymmetry between leading and non-leading
particles, which is negligible in NLO QCD, has been shown to increase with $\xF$
\cite{WA82,E769,E791}.
Also the momentum spectra of produced D mesons are harder than the NLO QCD
c quark predictions, especially for leading particles, see e.g. \cite{WA82,pertcharm2}.
These facts imply that nonperturbative effects are important in the
production of charmed hadrons. In the string fragmentation approach both these
aspects are included in the 'drag' effect, whereby a charm quark can
gain momentum when it is connected to a beam remnant. The extreme case of this
effect is the collapse of a small string into a single hadron, which gives rise
to a dependence on the flavour contents of the beam.

\section*{Basics of string fragmentation}

The Lund String Fragmentation Model \cite{AGIS} is best explored in $\ee$
annihilation, where the produced $\q$ and $\qbar$ are connected
by a linear force field with a string-like topology. The $\q \qbar$ production
process is described by perturbative QCD, with a parton-shower
approximation to higher orders. Radiated gluons are interpreted as 'kinks'
on the string. The nonperturbative hadronization of a string proceeds
via the production of $\q \qbar$ pairs in the colour force field, which
arrange themselves to produce the observed hadrons. A strongly
constrained fragmentation function can be derived from very general and
physically intuitive assumptions about the fragmentation process \cite{AGS}.
Because of the non-negligible mass of the charm quark the fragmentation
function has to be modified for heavy-flavour production
\cite{Bow}. This model has been implemented in the Monte Carlo program
\Py \cite{Pythia}, which has been tuned to $\ee$ experiments to give
a good description of available data.

\subsection*{Hadron-hadron collisions}

When we carry this model over to hadron-hadron physics, we again divide the
process into a perturbative and a nonperturbative part and assume
factorization between the two. In addition we assume that the
fragmentation process is universal, i.e.
the hadronization of a colour singlet is independent of how it was produced.
In a hadron-hadron collision, such as $\pi^- \p$, several ambiguities not present
in $\ee$ annihilation are introduced. The main ones are presented in
the following.

{\it Structure of the incoming hadrons.}
The particles that participate
in the collision are not point-like but have an internal structure.
The longitudinal structure is parameterized by parton distribution
functions (PDFs) which have been determined from other experiments such as deep
inelastic scattering (DIS). These will not be discussed in the following, but
in principle they give rise to some ambiguity, especially for small momentum
fractions.

{\it Structure of the beam remnant.}
When a parton has been picked out of a hadron, what is left continues in
the direction of the beam and is called a beam remnant. If the beam remnant
can be viewed as
consisting of two or more objects its structure must be described.
How this should be done is not known from first principles and
has not been studied much. This aspect is therefore parameterized in beam
remnant distribution functions (BRDFs) and several variants are considered.

{\it Primordial transverse momentum.}
The partons inside the hadron are confined
to a transverse dimension less than 1 fm; therefore by the uncertainty
principle the spread of the transverse momentum should be of the
order .2 GeV. This is modeled by adding a {\it primordial} $\kt$ to the partons
going into the hard scattering process. In the default version of \Py, $\kt$
is assumed distributed as a Gaussian with a width of 0.44 GeV. Several studies
\cite{pertcharm2,cprod,E791corr}
imply that this value is too small and a value of 1 GeV or more is needed to
describe the data. This remains somewhat of a mystery and will not be
resolved here.

{\it Small strings in hadronization. Quark masses.}
When the colour topology of an event has been determined, every colour singlet
is hadronized as a string would in $\ee$. This works for strings with a
mass of a few GeV or more. For strings with masses near the two-particle
threshold, the standard string fragmentation approach can not be used and some other
scheme is needed. As we will see later this introduces a large
dependence on the quark masses.

\subsection*{String topologies and the 'beam drag' effect}

\begin{figure}
\begin{center}
\begin{picture}(421,200)(0,0)
\SetOffset(0,100)
\Text(11,28)[r]{$\p$}
\Text(15,80)[r]{$\pi^-$}
\Line(15,30)(25,30)
\Line(15,80)(25,80)
\GOval(30,30)(20,5)(0){0.5}
\GOval(30,80)(20,5)(0){0.5}
\Text(48,40)[b]{$\u$}
\Text(48,70)[t]{$\ubar$}
\Line(35,40)(58,55)
\Line(35,70)(58,55)
\Gluon(58,55)(86,55){3}{4}
\Line(86,55)(110,40)
\Line(86,55)(110,70)
\Text(114,40)[l]{$\c$}
\Text(114,70)[l]{$\cbar$}
\Line(35,20)(106,20)
\Line(35,90)(110,90)
\Text(110,20)[l]{$\u\d$}
\Text(114,90)[l]{$\d$}
\put(124,130){\oval(4,20)[r]}
\put(123,180){\oval(4,20)[r]}

\SetOffset(141,100)
\Text(11,28)[r]{$\p$}
\Text(15,80)[r]{$\pi^-$}
\Line(15,30)(25,30)
\Line(15,80)(25,80)
\GOval(30,30)(20,5)(0){0.5}
\GOval(30,80)(20,5)(0){0.5}
\Gluon(35,40)(58,55){3}{3}
\Gluon(35,70)(58,55){3}{3}
\Gluon(58,55)(86,55){3}{4}
\Line(86,55)(110,40)
\Line(86,55)(110,70)
\Text(114,40)[l]{$\cbar$}
\Text(114,70)[l]{$\c$}
\Line(35,20)(106,20)
\Line(35,90)(106,90)
\Text(110,20)[l]{$\u\d$}
\Text(114,90)[l]{$\d$}
\Line(35,30)(106,30)
\Line(35,80)(106,80)
\Text(114,30)[l]{$\u$}
\Text(114,80)[l]{$\ubar$}
\put(265,135){\oval(4,11)[r]}
\put(265,155){\oval(12,70)[r]}
\put(265,175){\oval(4,11)[r]}

\SetOffset(287,100)
\Text(11,28)[r]{$\p$}
\Text(15,80)[r]{$\pi^-$}
\Line(15,30)(25,30)
\Line(15,80)(25,80)
\GOval(30,30)(20,5)(0){0.5}
\GOval(30,80)(20,5)(0){0.5}
\Gluon(35,40)(58,55){3}{3}
\Gluon(35,70)(58,55){3}{3}
\Gluon(58,55)(86,55){3}{4}
\Line(86,55)(110,40)
\Line(86,55)(110,70)
\Text(114,40)[l]{$\cbar$}
\Text(114,70)[l]{$\c$}
\Line(35,20)(106,20)
\Line(35,90)(106,90)
\Text(110,20)[l]{$\u\d$}
\Text(114,90)[l]{$\d$}
\Line(35,30)(106,30)
\Line(35,80)(106,80)
\Text(114,30)[l]{$\u$}
\Text(114,80)[l]{$\ubar$}
\put(411,145){\oval(14,50)[r]}
\put(411,155){\oval(9,50)[r]}
\put(411,165){\oval(4,50)[r]}

\SetOffset(0,10)
\Text(75,5)[]{(a)}
\ArrowLine(5,50)(65,50)
\Text(5,59)[l]{$\pi^-$}
\ArrowLine(130,50)(65,50)
\Text(130,45)[r]{$\p$}
\SetOffset(-10,10)
\LongArrow(75,50)(110,55)
\Text(115,57)[l]{$\d$}
\LongArrow(75,50)(100,75)
\Text(105,79)[l]{$\cbar$}
\LongArrow(75,50)(35,40)
\Text(33,35)[r]{$\u\d$}
\LongArrow(75,50)(65,30)
\Text(65,27)[t]{$\c$}
\DashLine(110,55)(100,75){3}
\DashLine(35,40)(65,30){4}

\SetOffset(140,10)
\Text(75,5)[]{(b)}
\ArrowLine(10,50)(75,50)
\Text(10,59)[l]{$\pi^-$}
\ArrowLine(130,50)(75,50)
\Text(130,45)[r]{$\p$}
\LongArrow(75,50)(110,55)
\Text(115,57)[l]{$\ubar$}
\LongArrow(75,50)(100,70)
\Text(105,74)[l]{$\c$}
\LongArrow(75,50)(35,40)
\Text(33,37)[t]{$\u$}
\LongArrow(75,50)(65,35)
\Text(64,29)[]{$\cbar$}
\LongArrow(75,50)(40,55)
\Text(40,64)[]{$\u\d$}
\LongArrow(75,50)(110,40)
\Text(112,34)[l]{$\d$}

\DashLine(110,55)(100,70){3}
\DashLine(35,40)(65,35){4}
\DashLine(40,55)(110,40){4}

\SetOffset(280,10)

\Text(75,5)[]{(c)}
\ArrowLine(10,50)(75,50)
\Text(10,59)[l]{$\pi^-$}
\ArrowLine(135,50)(75,50)
\Text(135,45)[r]{$\p$}
\LongArrow(75,50)(110,55)
\Text(115,57)[l]{$\ubar$}
\LongArrow(75,50)(100,70)
\Text(105,74)[l]{$\c$}
\LongArrow(75,50)(35,40)
\Text(33,37)[t]{$\u$}
\LongArrow(75,50)(65,35)
\Text(64,29)[]{$\cbar$}
\LongArrow(75,50)(40,55)
\Text(40,64)[]{$\u\d$}
\LongArrow(75,50)(110,40)
\Text(112,34)[l]{$\d$}

\DashLine(65,35)(110,40){3}
\DashLine(35,40)(110,55){4}
\DashLine(40,55)(100,70){4}
\end{picture}
\end{center}
\caption[]{Examples of different string configurations in a $\pi^-\p$ 
collision: (a) $\u\ubar \to \c\cbar$ has a unique colour flow; (b,c)
$\g\g \to \c\cbar$ with the two possible colour flows. Dashed lines are strings.}
\label{fig.strings}
\end{figure}
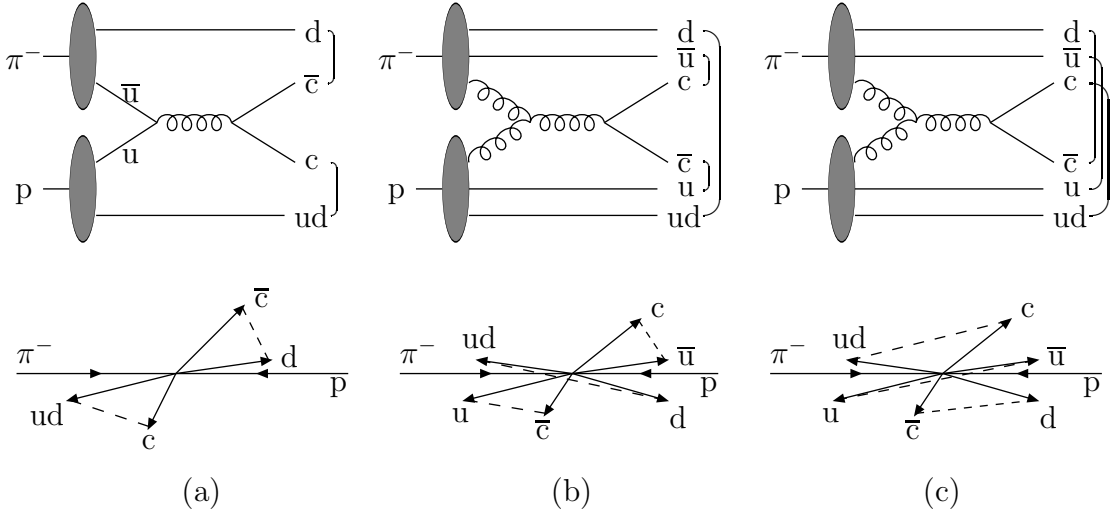

In a $\pi^- \p$ collision we include charm production via the leading-order
production mechanisms of quark and gluon fusion ($\q\qbar \to \c\cbar$ and
$\g\g \to \c\cbar$ respectively).
The partons of the hard interaction and of the beam remnants 
are connected by strings, representing the confining colour field
\cite{AGIS}. Each string contains a colour triplet endpoint, a number
(possibly zero) of intermediate gluons and a colour anti-triplet end.
The string topology can usually be derived from the colour flow of the
hard process. For instance, consider the process $\u\ubar \to \c\cbar$
in a $\pi^-\p$ collision. The colour of the incoming $\u$ is inherited
by the outgoing $\c$, so it will form a colour-singlet together with the
proton remnant, here represented by a colour anti-triplet $\u\d$ diquark. 
In total, the event will thus contain two strings, one $\c$--$\u\d$ and
one $\cbar$--$\d$ (Fig.~\ref{fig.strings}a). In $\g\g \to \c\cbar$
a similar inspection shows that two distinct colour topologies are 
possible. Representing the 
proton remnant by a $\u$ quark and a $\u\d$ diquark (alternatively $\d$ 
plus $\u\u$), one possibility is to have three strings $\c$--$\ubar$, 
$\cbar$--$\u$ and $\d$--$\u\d$ (Fig.~\ref{fig.strings}b), and the other
is the three strings $\c$--$\u\d$, $\cbar$--$\d$ and $\u$--$\ubar$
(Fig.~\ref{fig.strings}c).

Other production mechanisms such as charm excitation and charm
from parton showers (i.e. higher order effects) are not included in this
study. Because of the relatively low virtuality of the hard process at fixed
target energies the second mechanism is negligible, but it will become increasingly
important at higher energies; at the LHC, e.g., this mechanism will dominate
over the fusion mechanisms. Charm excited from the sea could give a
non-negligible contribution, but we will not include it here.

Consider a colour singlet in Fig.~\ref{fig.strings} containing a charm endpoint.
The hadronization of this string is performed in the CM system of the string.
In that frame hadronization always results in a deceleration of the quark.
After the rotation and boost to the hadron-hadron CM system, on the other hand,
the net result of hadronization can be either an acceleration or a deceleration
of the charm quark. This is interpreted as the beam remnant dragging the charm quark
in the direction of the beam. This effect alone does not account for the asymmetry
because of the cancellation between the diagrams in Fig~\ref{fig.strings}b
and c. We must therefore consider the flavour contents of the beam.

\subsection*{Cluster collapse of small strings}

In $\ee$ annihilation the string mass is fixed by the CM energy
of the process. In a hadron-hadron collision, on the other hand, charmed
strings can have any mass ranging from $m_{\q}$+$m_{\c}$ to $\sqrt{s}$.
For string masses larger than some cut-off (here taken as $m_{\q}$+$m_{\c}$+$m_0$,
with $m_0$$\approx$1 GeV and q a light quark) the Lund string fragmentation approach can be used.
The model assumes a continuous final-state phase space, and an
iterative scheme is used which demands that at least two particles are
produced from a string. A string with a mass smaller than the cut-off we call a
{\it cluster} and it is hadronized in the following way:

{\it Cluster decay.} A $\q\qbar$ pair is created, using standard flavour
selection, from the force-field connecting the cluster endpoints, and two hadrons
are produced. If kinematically possible the cluster will decay isotropically
into these two hadrons.

{\it Cluster collapse.} If it is not kinematically possible for the
cluster to decay into two hadrons, it will be forced to collapse into a
single hadron under conservation of the flavour quantum numbers.
Since the mass of the cluster most likely will not correspond to any
physical state (e.g. D or $\D^*$), the energy and momentum of the cluster must
be slightly modified in order to put it on the hadronic mass shell.

There are two main ambiguities in this scheme. The first is that there is no clear
separation between the two hadronization mechanisms, and the second is
energy/momentum conservation in the cluster collapse.

To justify the cluster collapse approach we use an argument based on local duality,
which has been shown to hold in $\ee$ annihilation, DIS \cite{duality}
and $\tau$ decay \cite{taudecay}.
In $\ee$ annihilation into hadrons around the $\c\cbar$ threshold the observed
cross section consists of peaks at $\mathrm{J}/\psi$ and $\psi'$
and a continuum above the $\D\Dbar$ threshold. The perturbatively calculated
cross section, on the other hand, is continuous from $\sqrt{s}=2m_{\c}$ onwards.
However, if the experimental cross section is suitably smeared,
it approximately agrees with the perturbative one.
Another way of stating this is that the
integrated cross section (over $\sqrt{s}$) should be the same provided that the
integration interval is suitably large. We use the same argument in the present case,
by replacing $\sqrt{s}$ with $M_{\mathrm{string}}$ and the $\mathrm{J}/\psi$ and
$\psi'$ peaks with $\D$ and $\D^*$. The duality argument could then be stated in
the following way:

\begin{equation}
\int_{m_1}^{m_2}
\frac{\mathrm{d}\sigma_{\mathrm{Partons}}}{{\mathrm{d}}M_{\mathrm{String}}}
{\mathrm{d}}M \approx
\int_{m_1}^{m_2}
\frac{\mathrm{d}\sigma_{\mathrm{Hadrons}}}{{\mathrm{d}}M_{\mathrm{String}}}
{\mathrm{d}}M
\end{equation}

\begin{figure}
\begin{center}
\mbox{\epsfig{file=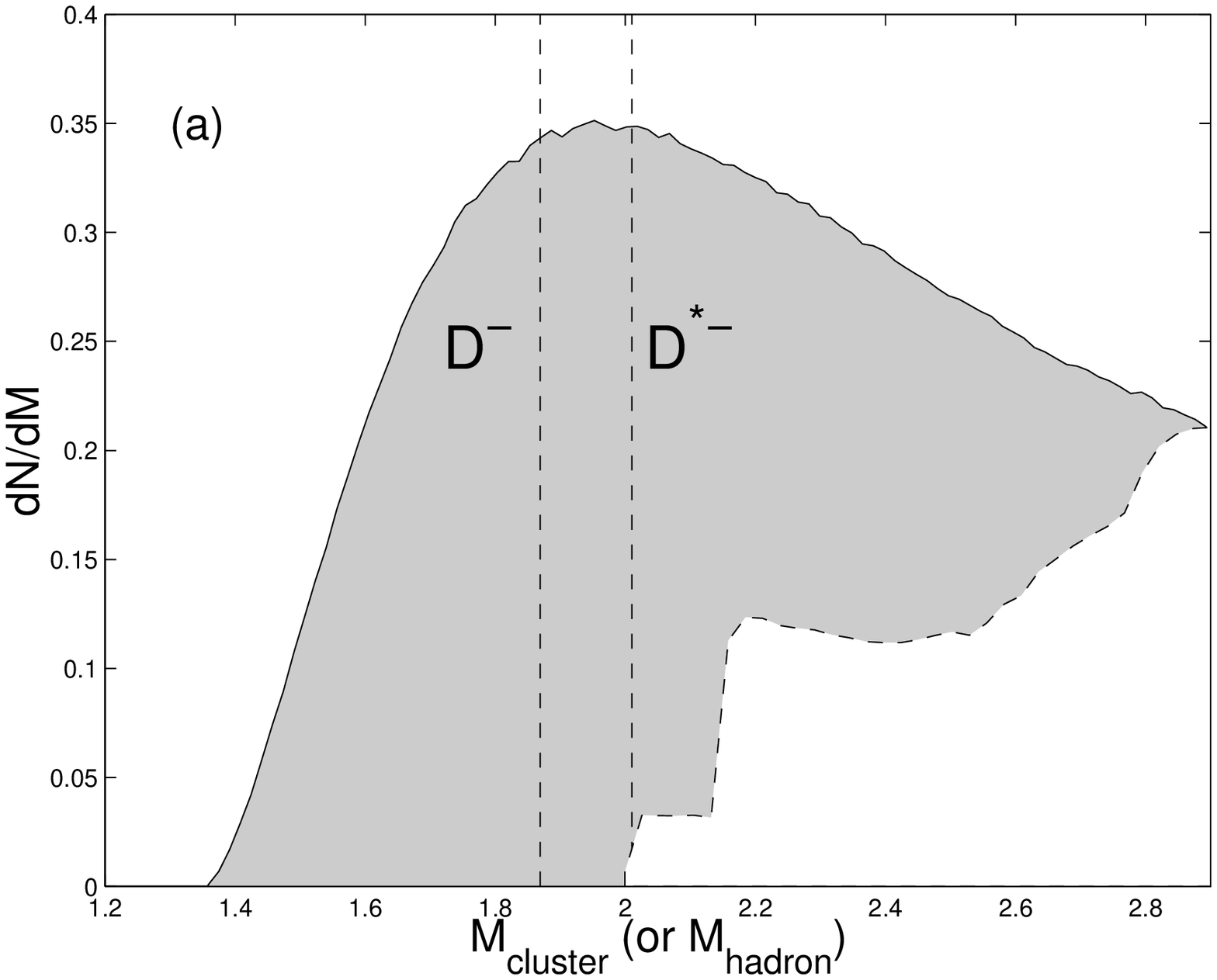,width=73mm}}
\mbox{\epsfig{file=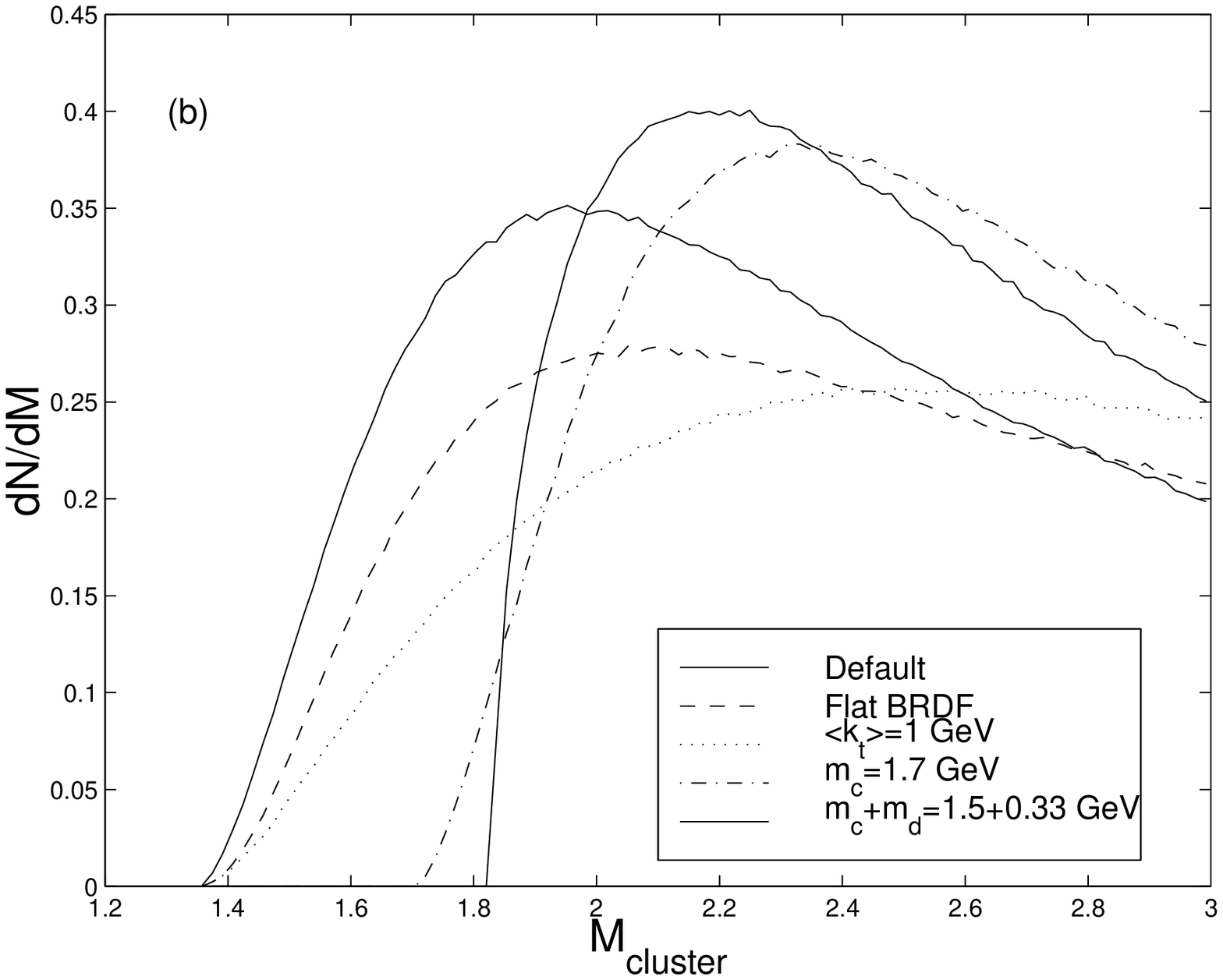,width=73mm}}
\end{center}
\caption[]{(a) Distribution of cluster (full) and meson (dashed) masses in the
string model. Clusters within the gray area collapse to $\D^-$ or $\D^{*-}$.
(b) Dependence of the parton level mass distribution on some parameters of the model.}
\label{fig.cmass}
\end{figure}

Fig.~\ref{fig.cmass}a shows how this looks using \Py~with standard parameters.
The solid line is the mass distribution of produced clusters at the parton level and
the dashed one is the produced hadrons. The clusters in the gray area have
collapsed into single hadrons. The parton level
distribution depends on many of the parameters such as the BRDF, primordial $\kt$,
and quark masses. This is shown in Fig~\ref{fig.cmass}b. Consider e.g. an
increase of the charm mass. The threshold will be shifted towards higher
values and fewer clusters will be forced to collapse (the gray area is decreased).
It is also possible to decrease the number of collapses from above by
increasing the probability for a cluster above the $\D \pi$ threshold to decay.

\begin{figure}[ht]
\begin{center}
\mbox{\epsfig{file=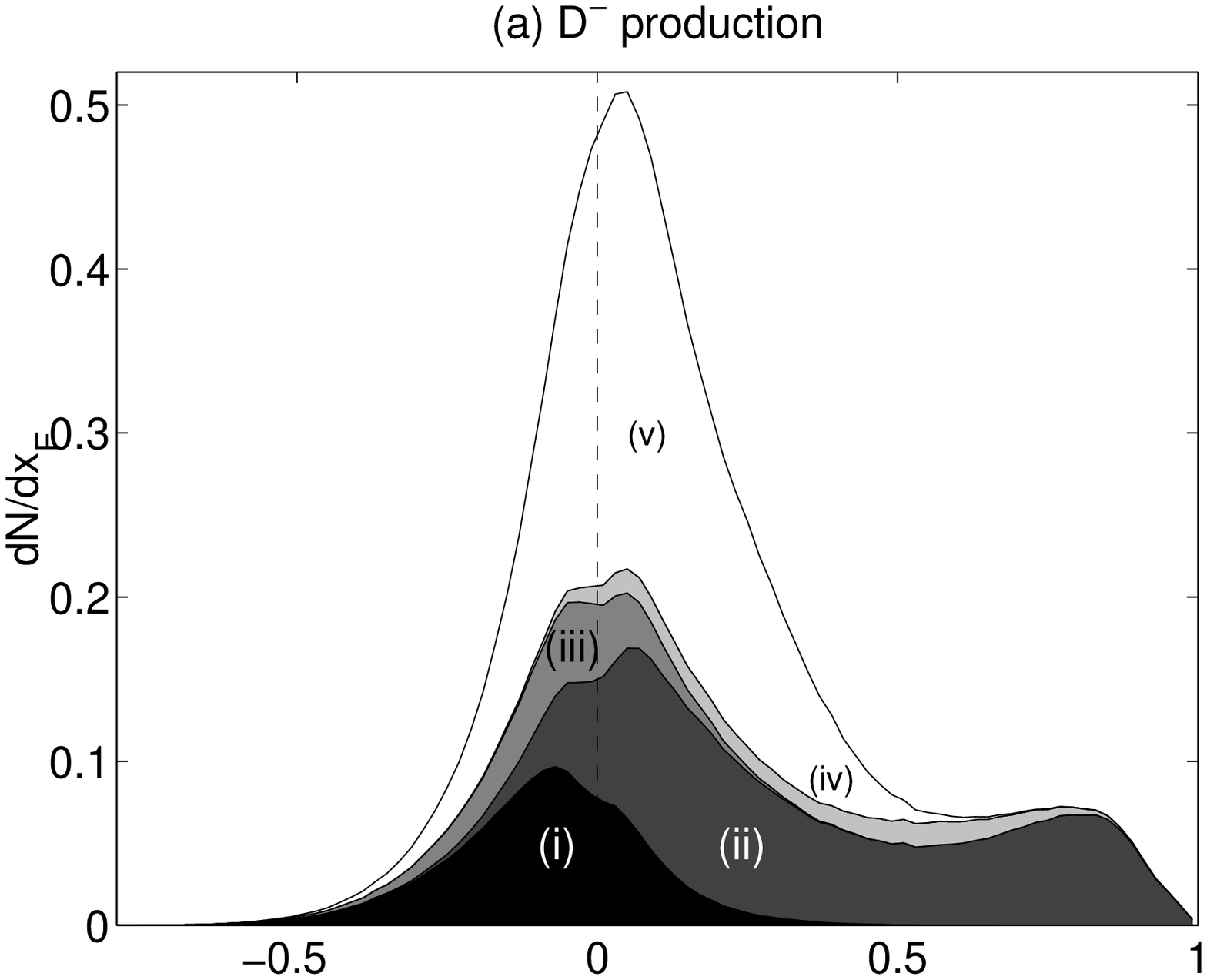, width=48mm}}
\mbox{\epsfig{file=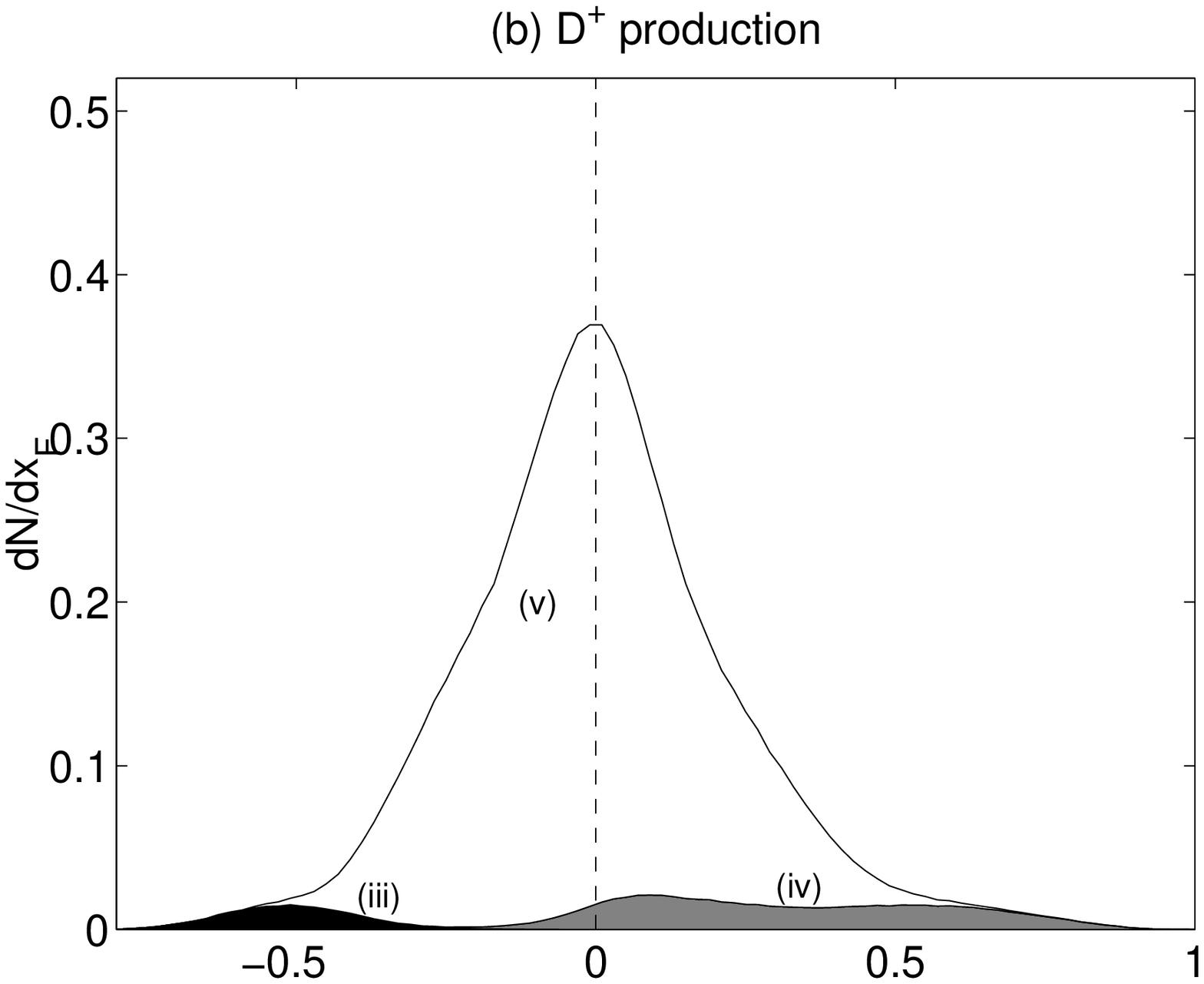, width=48mm}}
\mbox{\epsfig{file=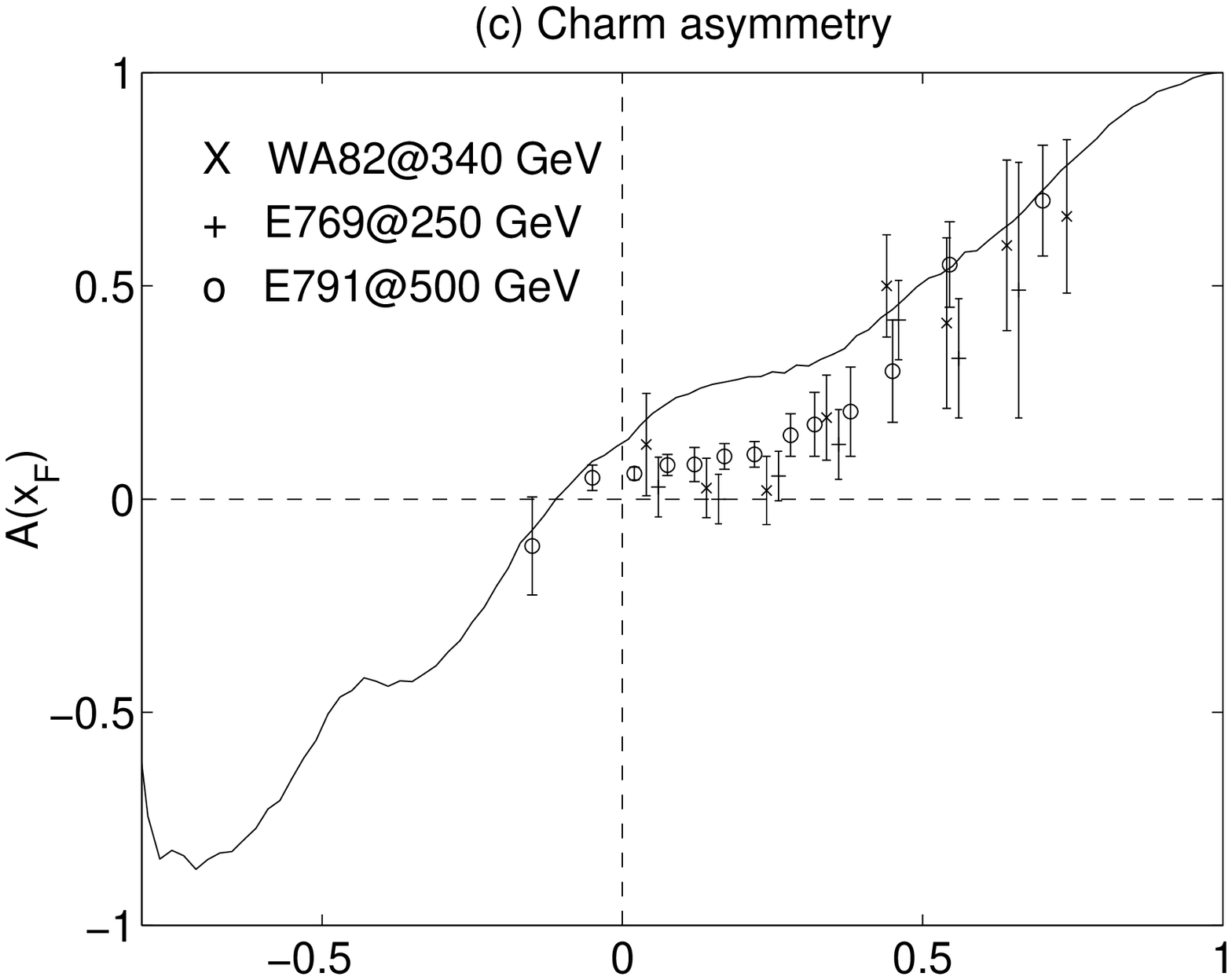, width=48mm}}
\mbox{\epsfig{file=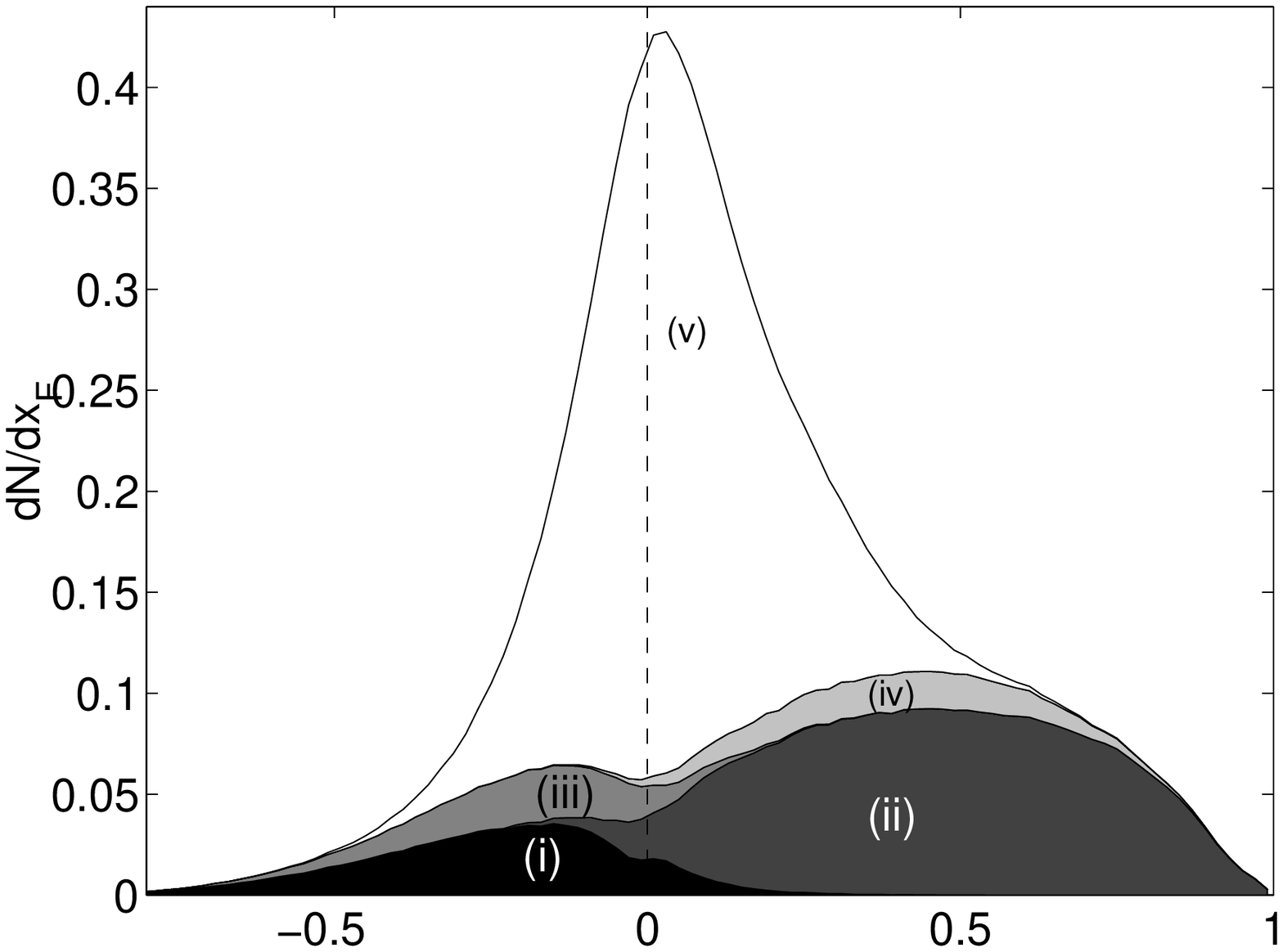, width=48mm}}
\mbox{\epsfig{file=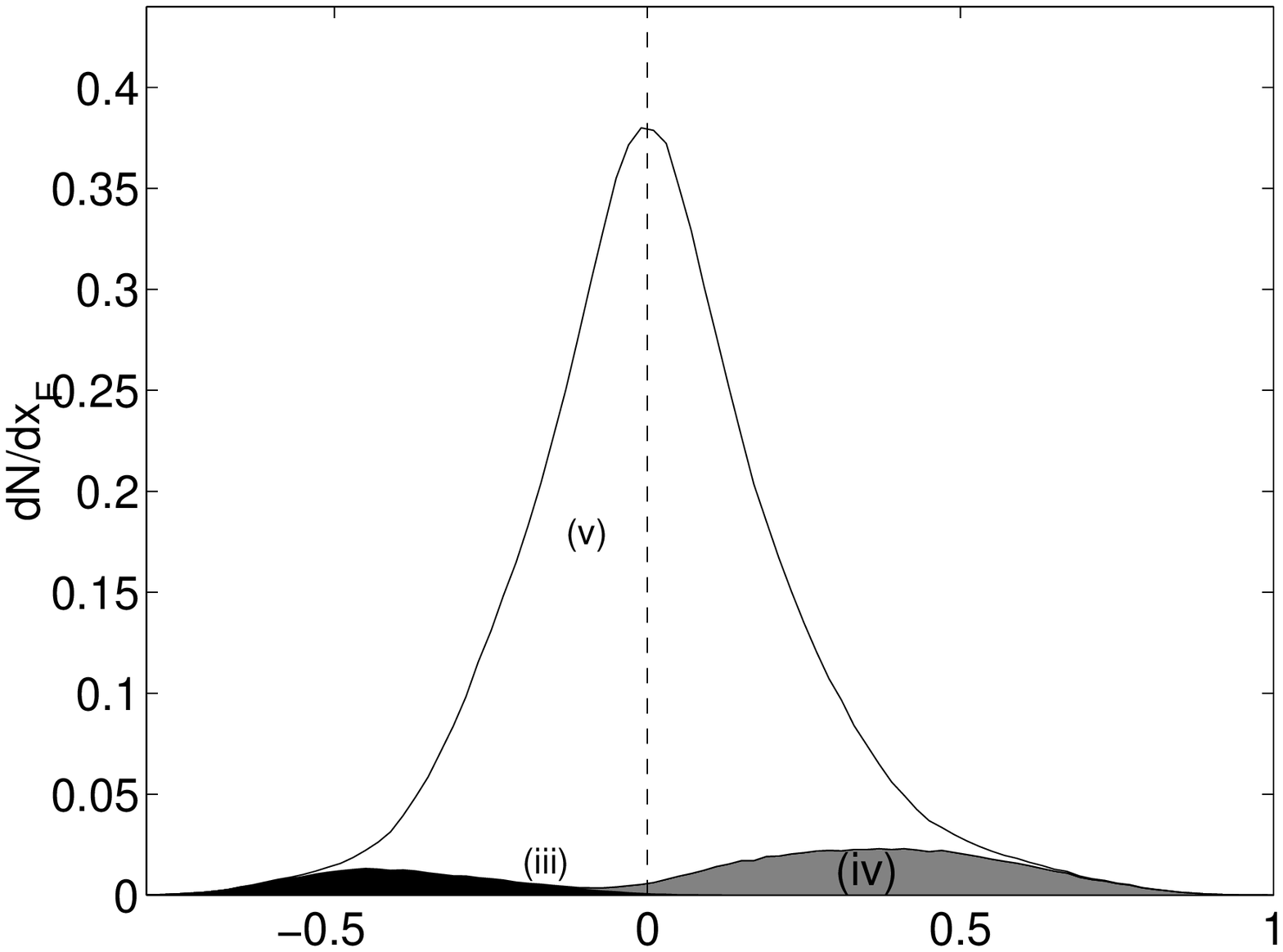, width=48mm}}
\mbox{\epsfig{file=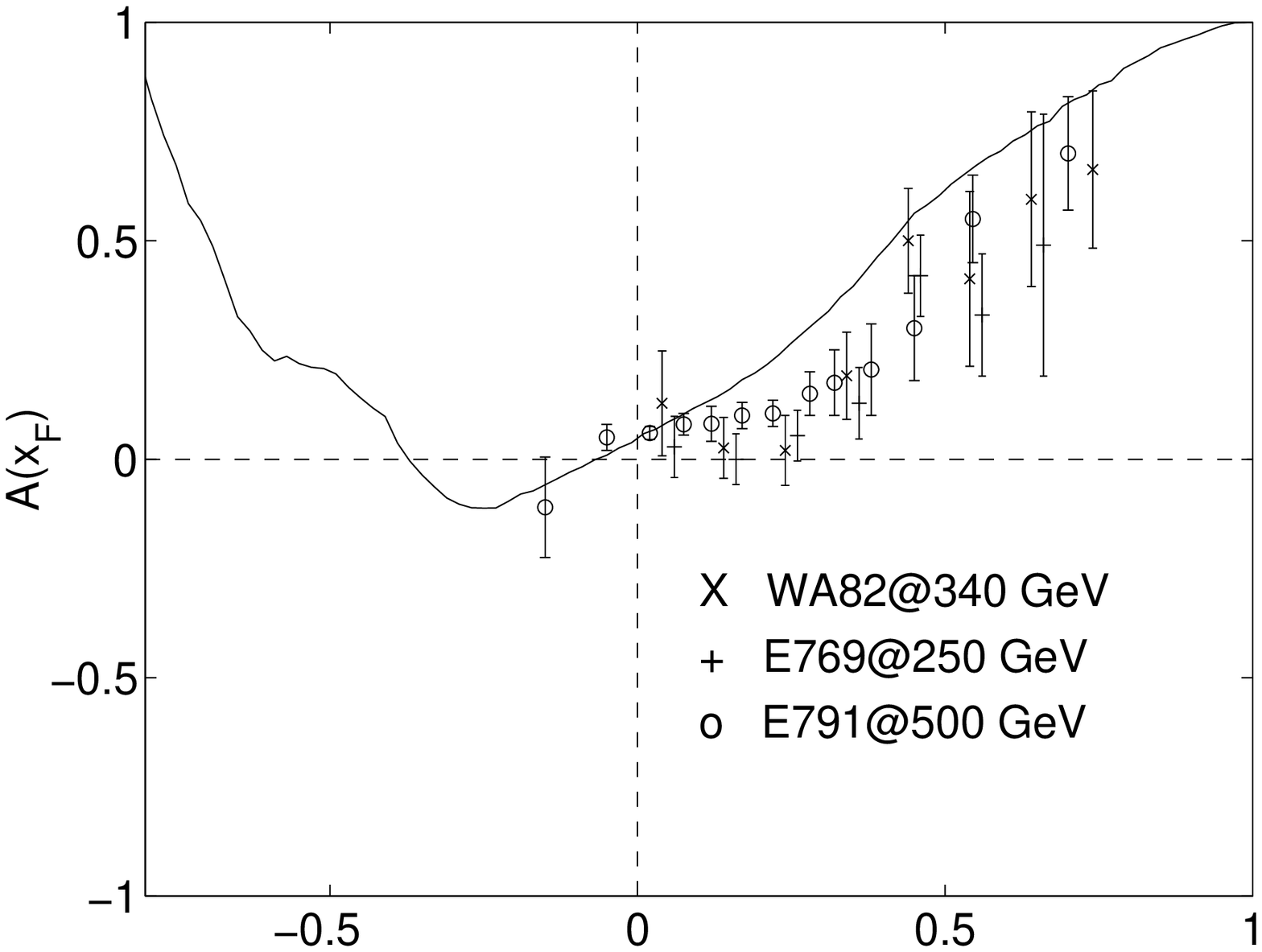, width=48mm}}
\mbox{\epsfig{file=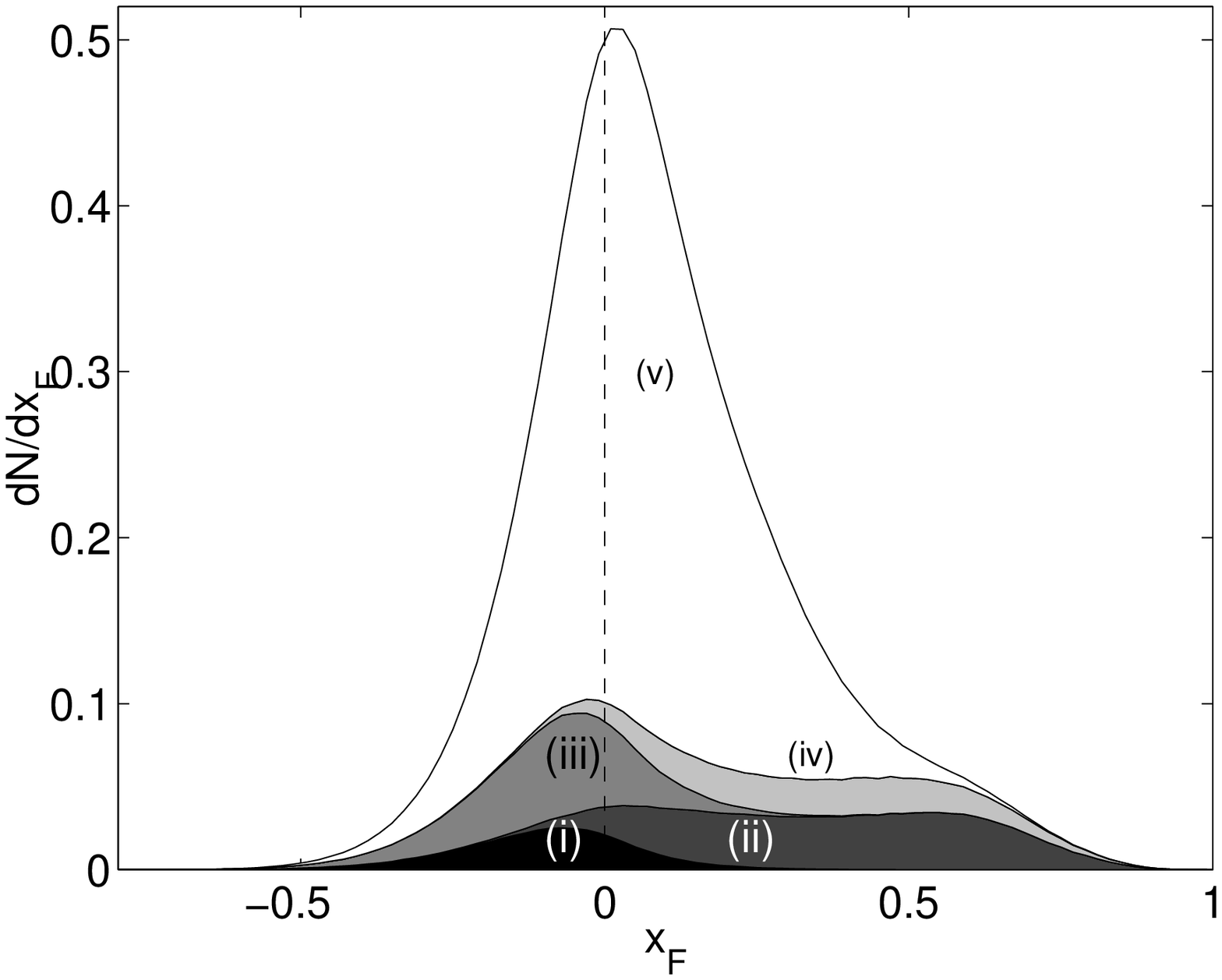, width=48mm}}
\mbox{\epsfig{file=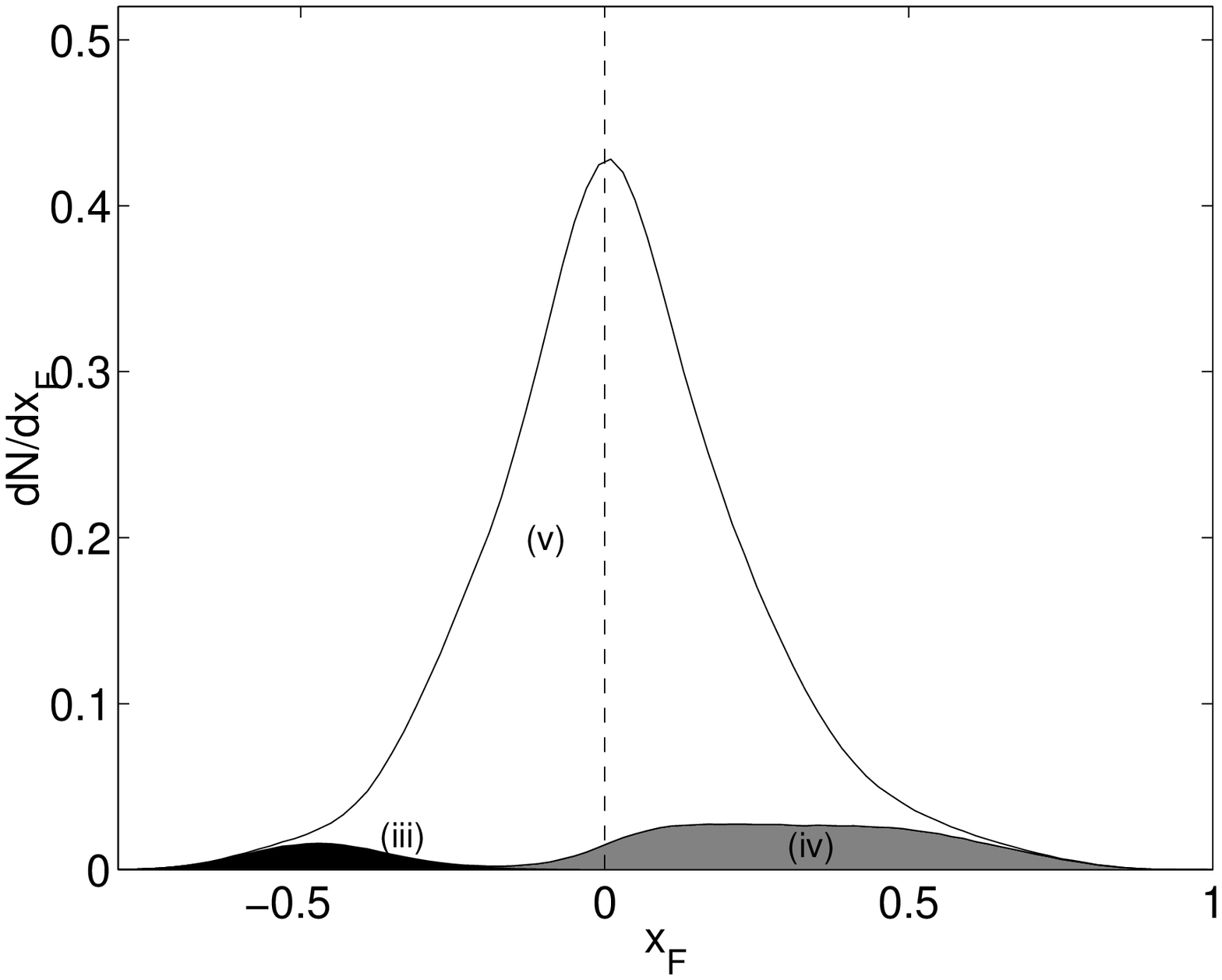, width=48mm}}
\mbox{\epsfig{file=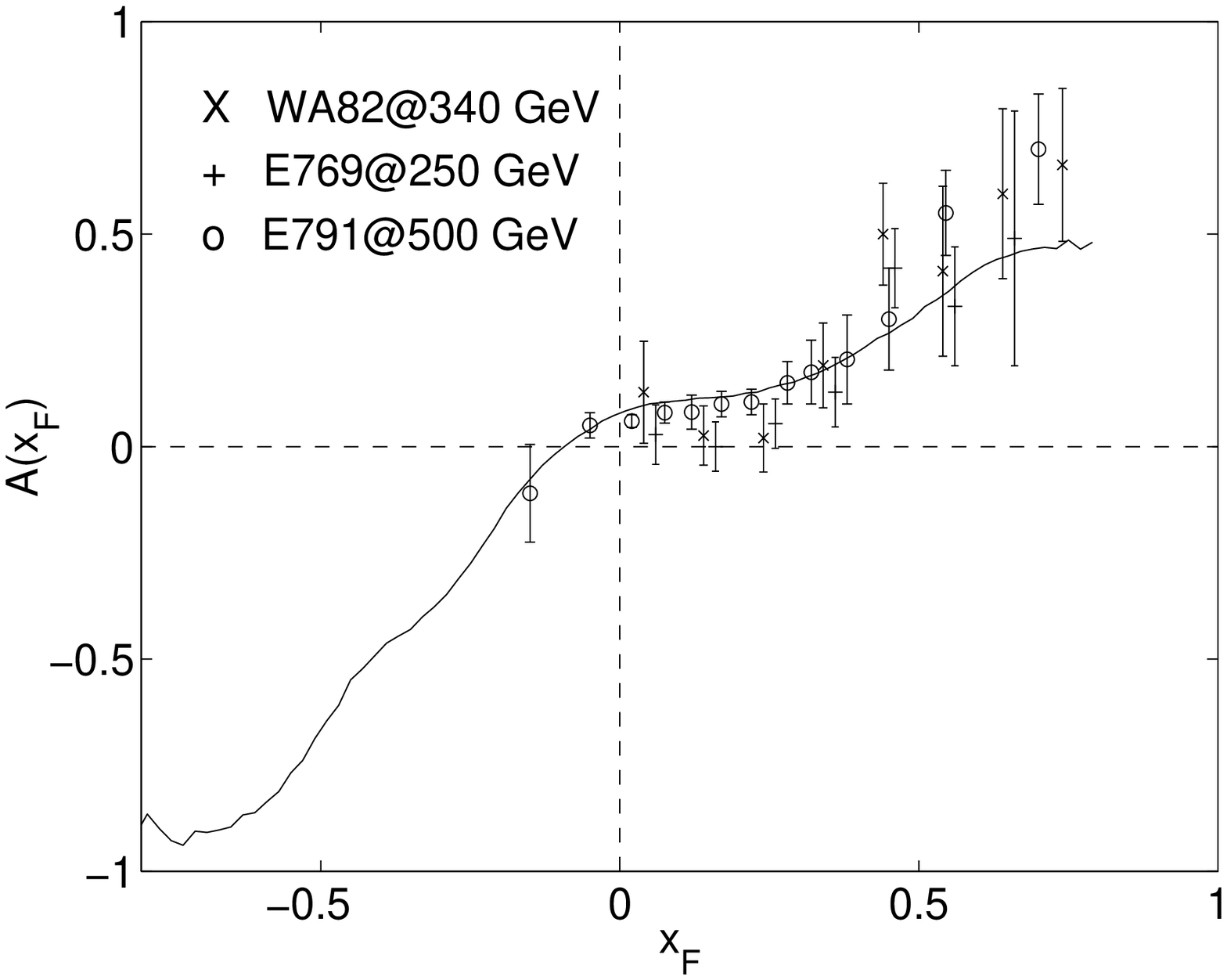, width=48mm}}
\end{center}
\caption[]{(a) $\D^-$ and (b) $\D^+$ meson production in a $\pi^-\p$ collision at a $\pi^-$
beam momentum of 500 GeV with different parameter sets. From top to bottom
these are: default, using flat BRDFs and the new parameter set presented in
this talk. The distributions are divided into different production channels:
(i) Cluster collapse, light quark from p end,
(ii) Cluster collapse, light quark from $\pi^-$ end,
(iii) Cluster decay, light quark from p end,
(iv) Cluster decay, light quark from $\pi^-$ end and
(v) String fragmentation.
(c) The resulting asymmetry.}
\label{fig.channels}
\end{figure}

Fig.~\ref{fig.channels} shows the $\xF$ distributions for different production
channels and different parameter sets. These parameters will be discussed in
the following. The explanation of the leading particle asymmetry
in this model is that $\D^+$ cannot be produced from cluster collapse because it
has no quark in common with the beam.

\subsection*{Dependence on parameters}

In this section the different parameters that have already been introduced
will be discussed in more detail. We start with the BRDFs.
How the energy and momentum in the beam remnant should be split between the constituents
is not known from first principles, so we consider two extreme cases. In the
first (default) scenario, we use BRDFs similar to those in PDFs, where one object
takes a small fraction of the available energy.
In the other extreme, we use naive counting rules where the energy is shared
democratically between the constituents. How this effects the distributions
is shown in Fig.~\ref{fig.channels}. Most notably the dip around $\xF \sim 0.5$ in the cluster
collapse distribution is smeared out when an even sharing is used. Also note
that the asymmetry is much more sensitive to the BRDF in the proton region,
where one of the constituents of the beam remnant is a diquark. In the tuned
version we will use an intermediate energy sharing.

We now come to the problem of energy/momentum conservation in the cluster collapse.
To understand the dependence on this we consider two mechanisms that
are of opposite nature:

{\it Old method: 'far away'.} In the first scheme, energy and momentum is
shuffled to the parton in the event that is farthest from the cluster, in order
to minimize the recoil. In this approach the D meson is
often harder than the cluster.

{\it New method: 'local'.} In this new scheme we conserve energy
locally by exchanging 'gluons' between the cluster and the string in the
event that is closest to the cluster.

The details are presented in \cite{cprod} and the conclusion is that the
observables are not sensitive to the details of the energy/momentum
conservation scheme, except for $\xF > 0.5$, where cluster collapse dominates
but data is scarce.

Looking at Fig.~\ref{fig.channels} we see that the reason for the large
asymmetry using the default parameters is that they allow many clusters
to collapse for $\xF > 0$. The following set of parameters are inspired
by the E791 collaboration and data from WA82 and they aim at decreasing
the asymmetry by decreasing the number of clusters that collapse into
one particle \cite{new}.

\begin{Itemize}
\item{Quark masses.} 
The charm mass used in \Py~is by default set to the current algebra one
(1.35 GeV). This is the value used in e.g. Higgs physics but it is
far from obvious that this is the relevant mass in the present case.
We therefore use constituent quark masses: $m_{\c}=$ 1.5 GeV;
$m_{\d}=m_{\u}=$ 0.33 GeV; $m_{\s}=0.5$ GeV. This will shift the threshold
of the distribution of cluster masses and thus decrease the number
of cluster collapses.
\item{Cluster decay.} Increase the probability for a cluster to decay.
\item{Cluster collapse.} Use new 'local' collapse mechanism.
\item{BRDF.} We use an intermediate energy sharing that is more democratic,
but not completely.
\item{Primordial $\kt$.} The width of the Gaussian primordial $\kt$ is
increased from 0.44 to 1.0 GeV. This allows cluster collapses between a
charm quark and a beam remnant to occur also at fairly large values of
$\pt$, thus leading to an essentially $\pt$-independent asymmetry. In addition,
the $\pt$ kick added to charm quarks and beam remnants tends to increase
the average invariant mass of the produced clusters, thereby reducing the
number of cluster collapses.
\end{Itemize}

\section*{Comparison with experiment}

\begin{figure}[t]
\begin{center}
\mbox{\epsfig{file=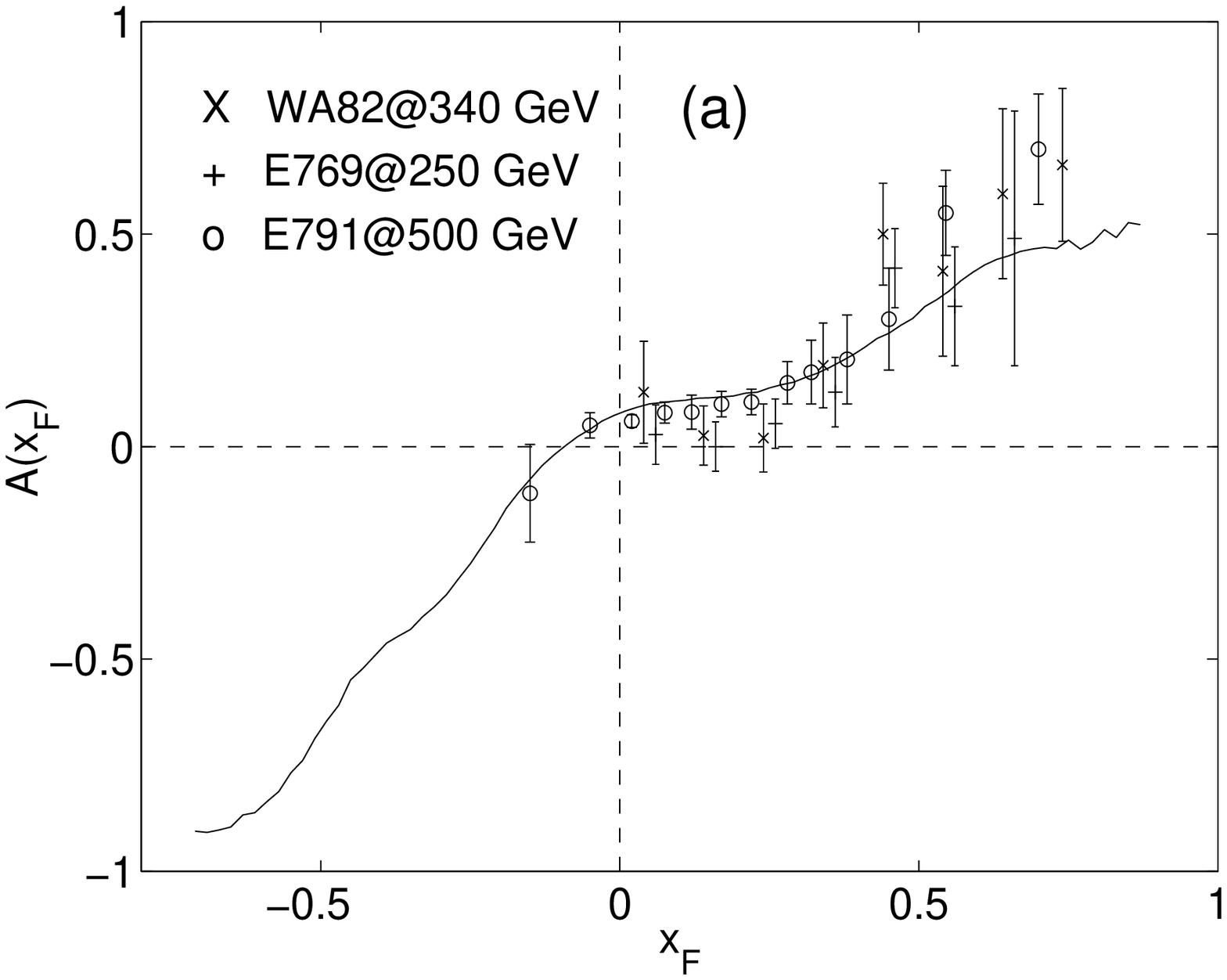, width=70mm}}
\mbox{\epsfig{file=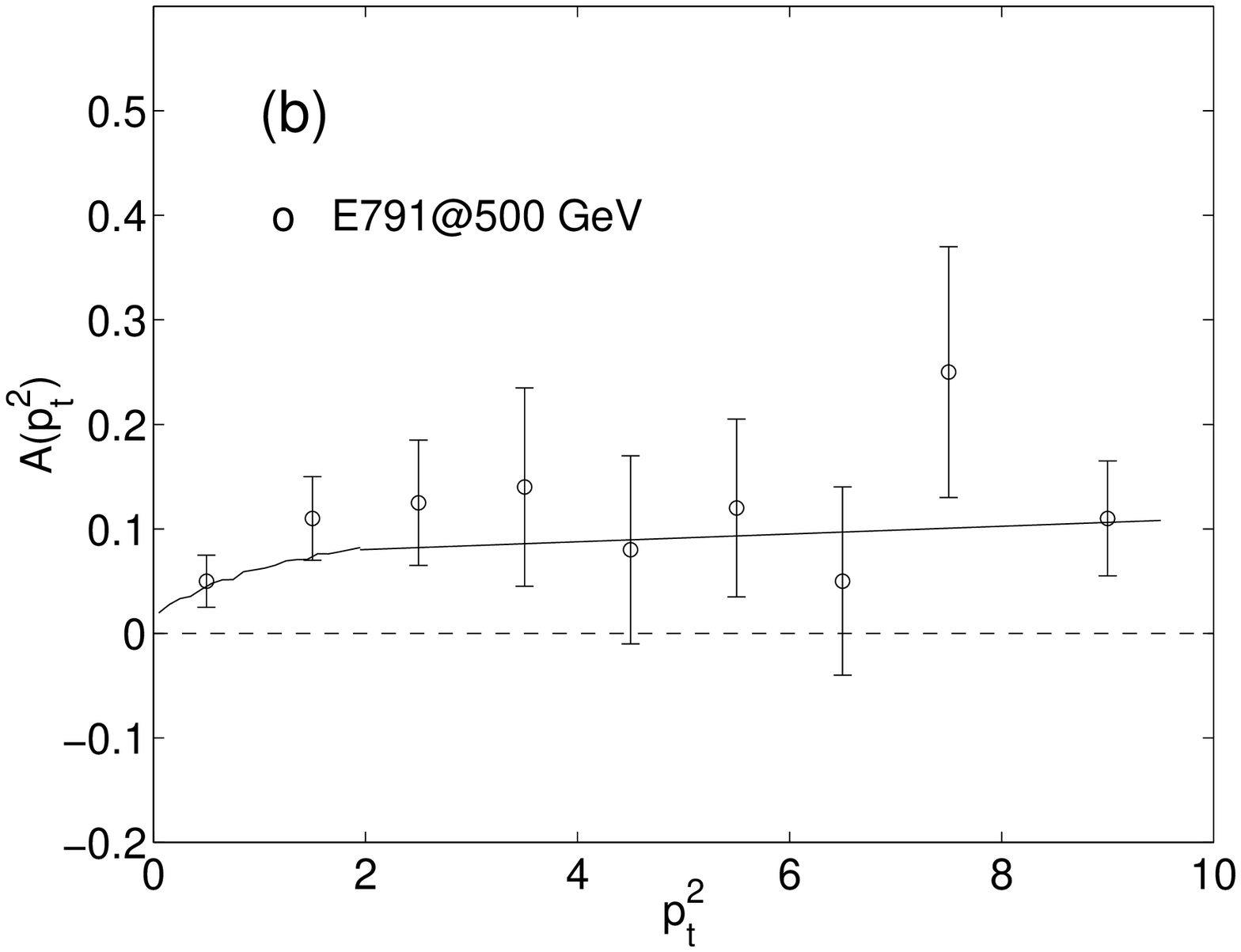, width=70mm}}
\mbox{\epsfig{file=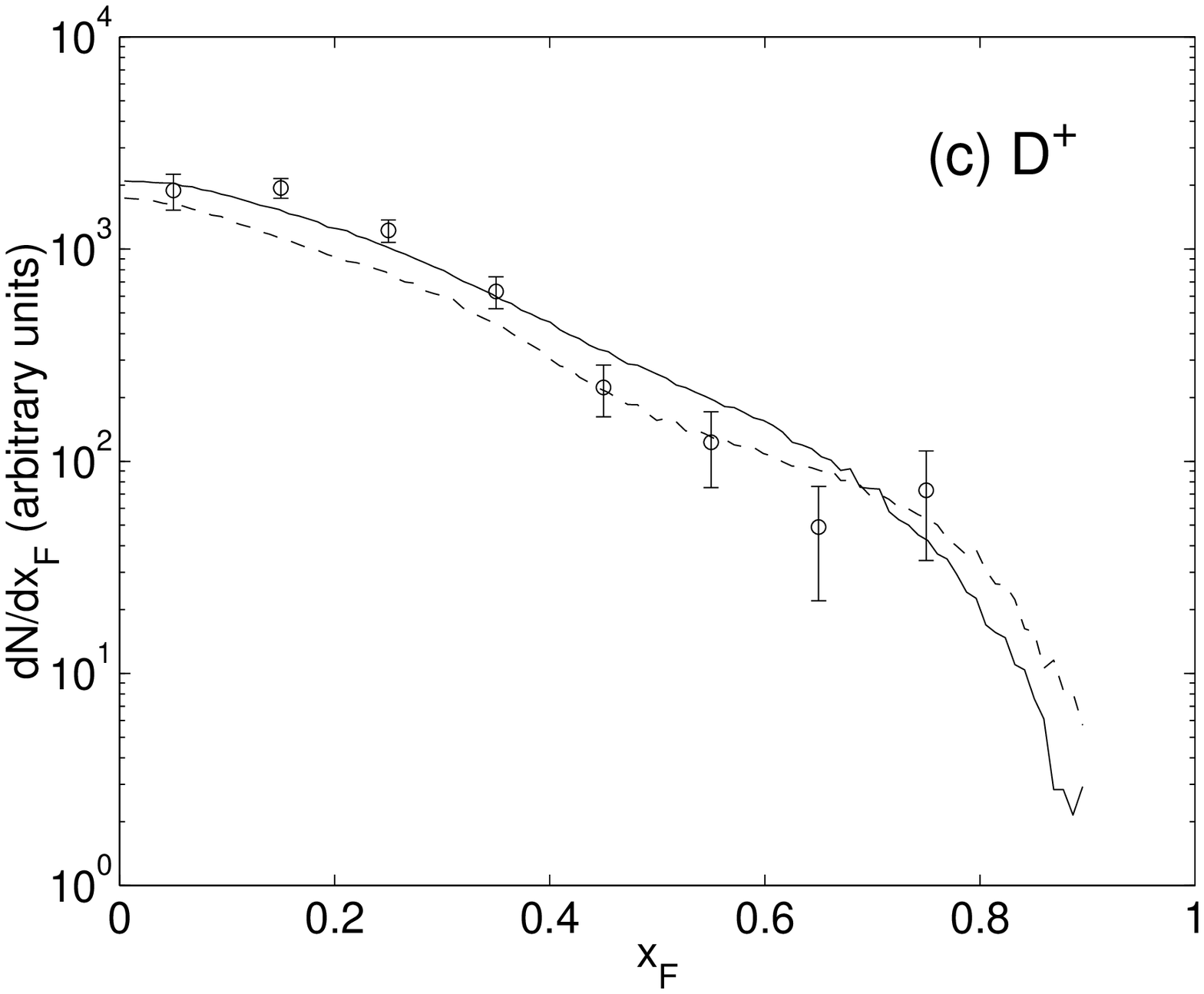, width=70mm}}
\mbox{\epsfig{file=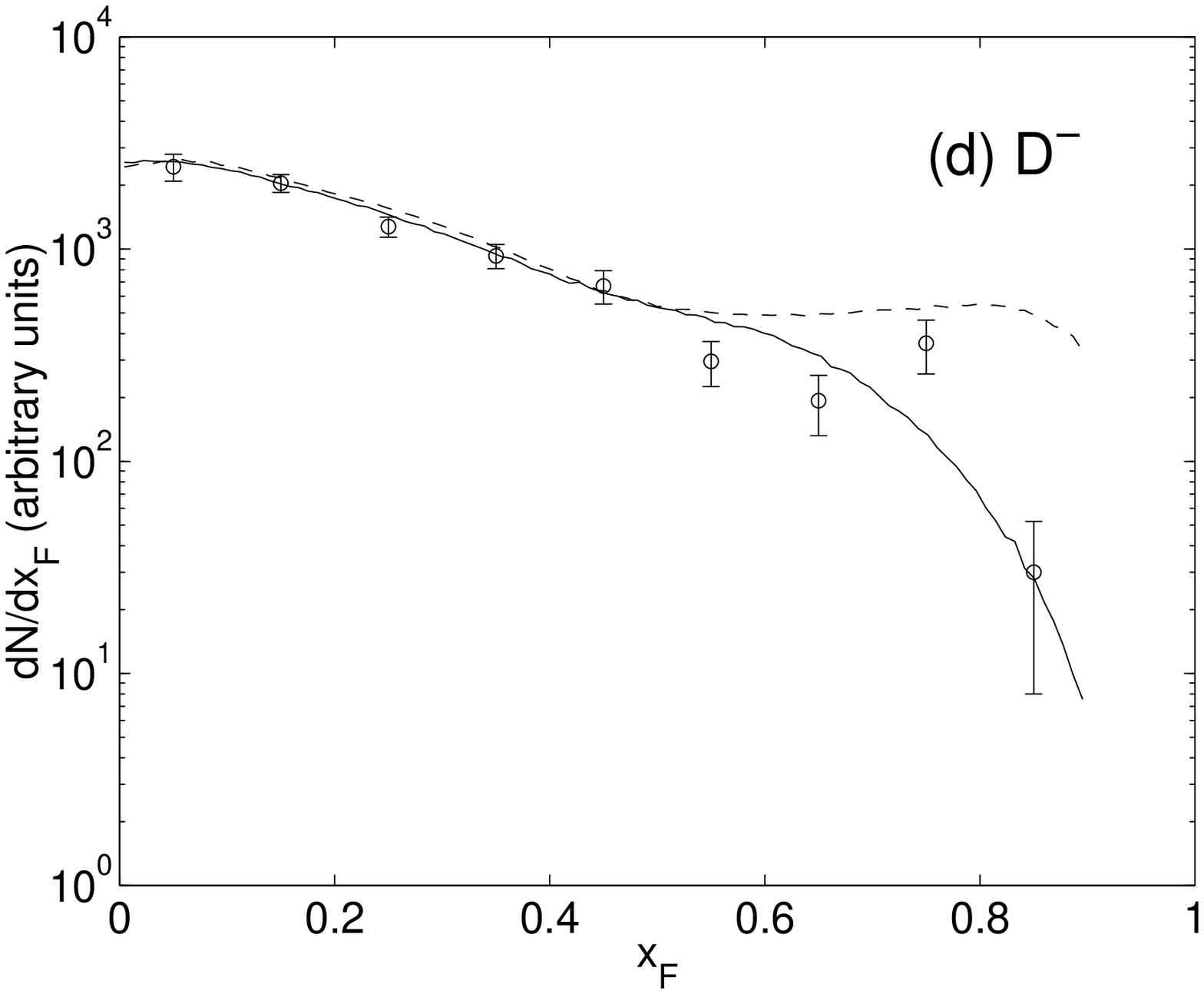, width=70mm}}
\end{center}
\caption[]{Comparison between the new \Py~parameter set and data.
Asymmetry as a function of (a) $\xF$ and (b) $\pt^2$. Single-charm spectra
from the WA82 CERN experiment \cite{WA82} for (c) $\D^+$ and (d) $\D^-$
(dashed line is standard parameters and the full line is the new).}
\label{fig.new} 
\end{figure}

In this section we compare the model (with the new parameter set \cite{new}) to some
data. Fig.~\ref{fig.new} shows the asymmetry as a function of $\xF$ and $\pt^2$
as well as single-charm spectra from WA82 for $\D^+$ and $\D^-$
individually. In this case the data is nicely described by the new
parameter set.

\begin{figure}
\begin{center}
\mbox{\epsfig{file=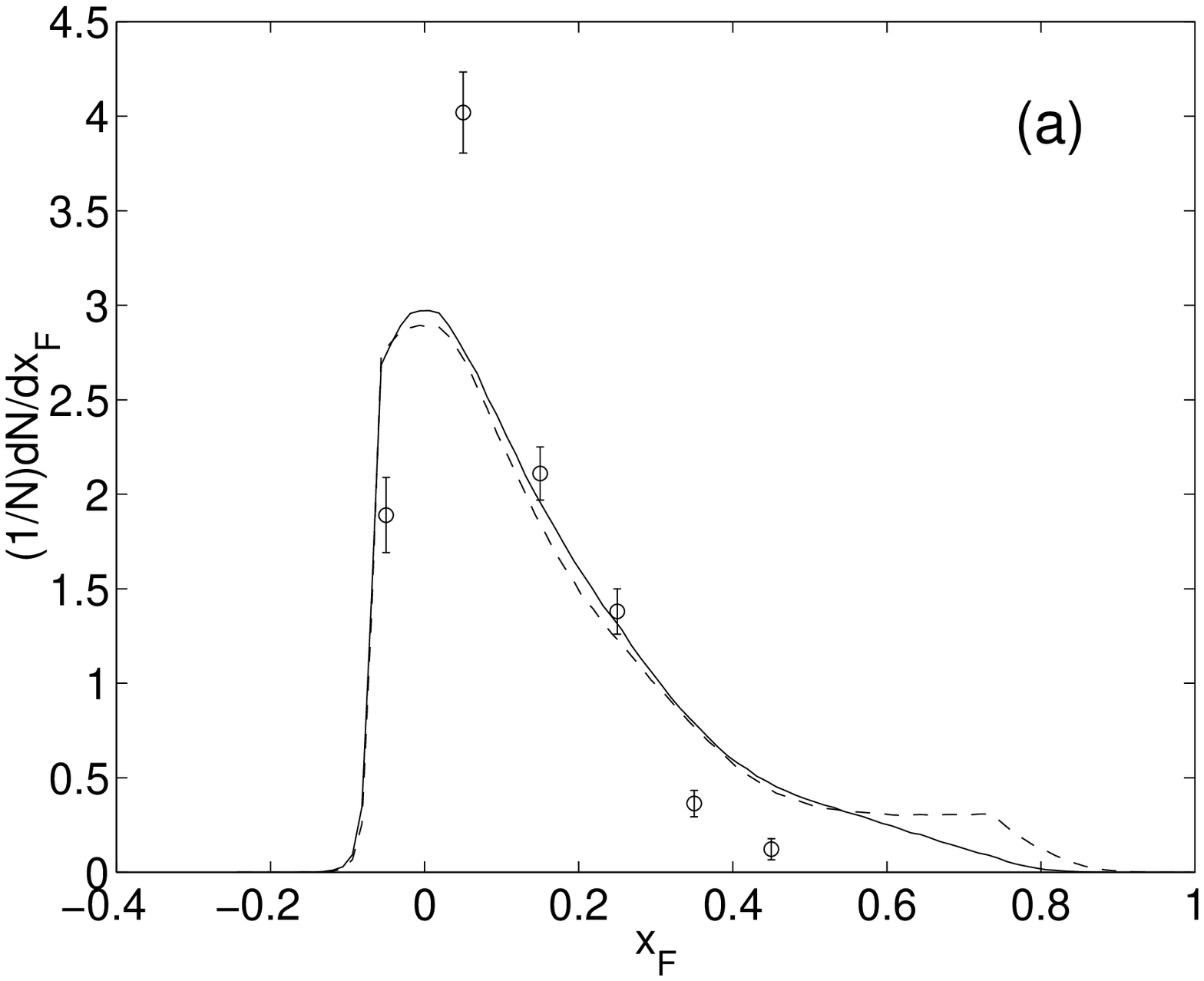, width=48mm}}
\mbox{\epsfig{file=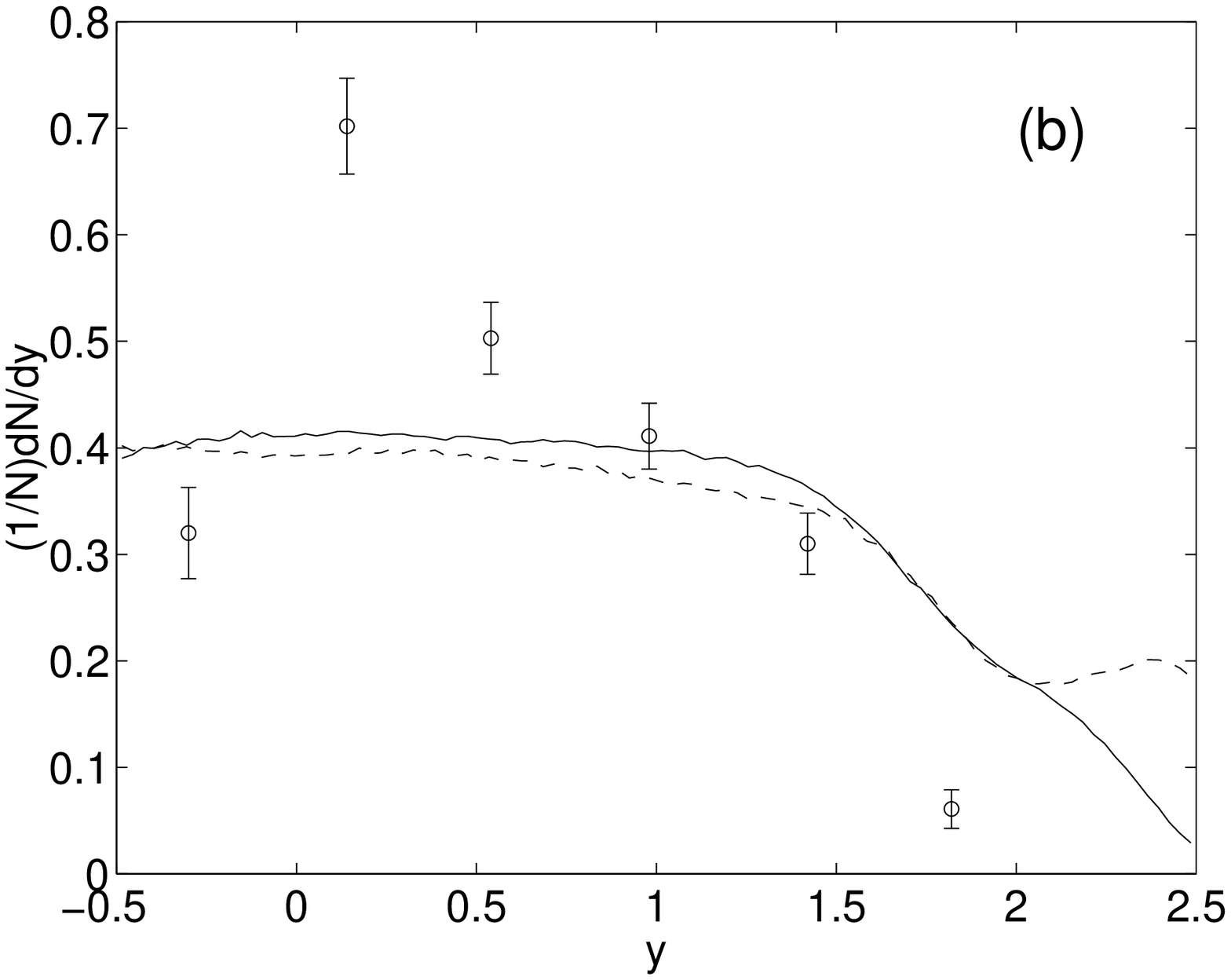, width=48mm}}
\mbox{\epsfig{file=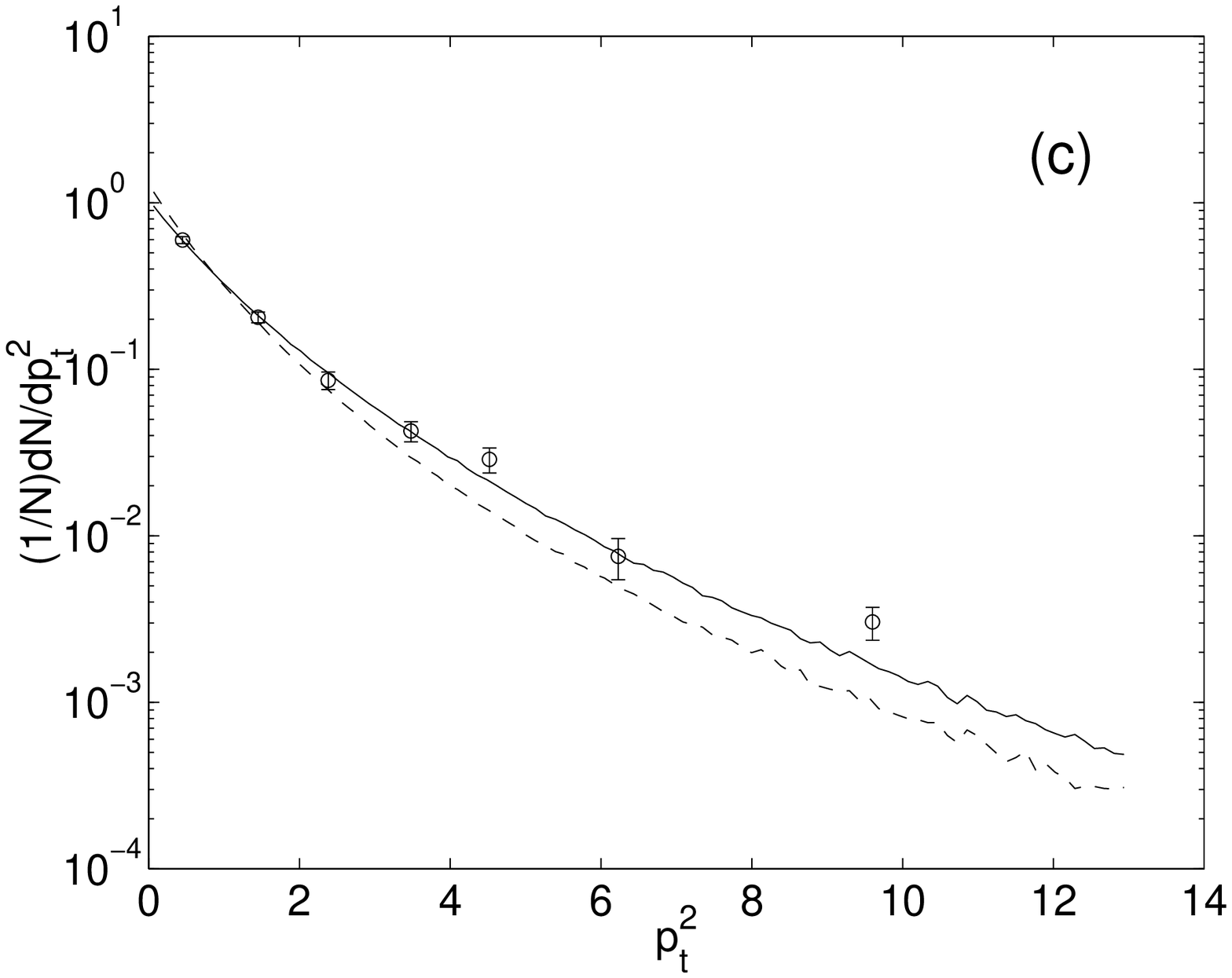, width=48mm}}
\mbox{\epsfig{file=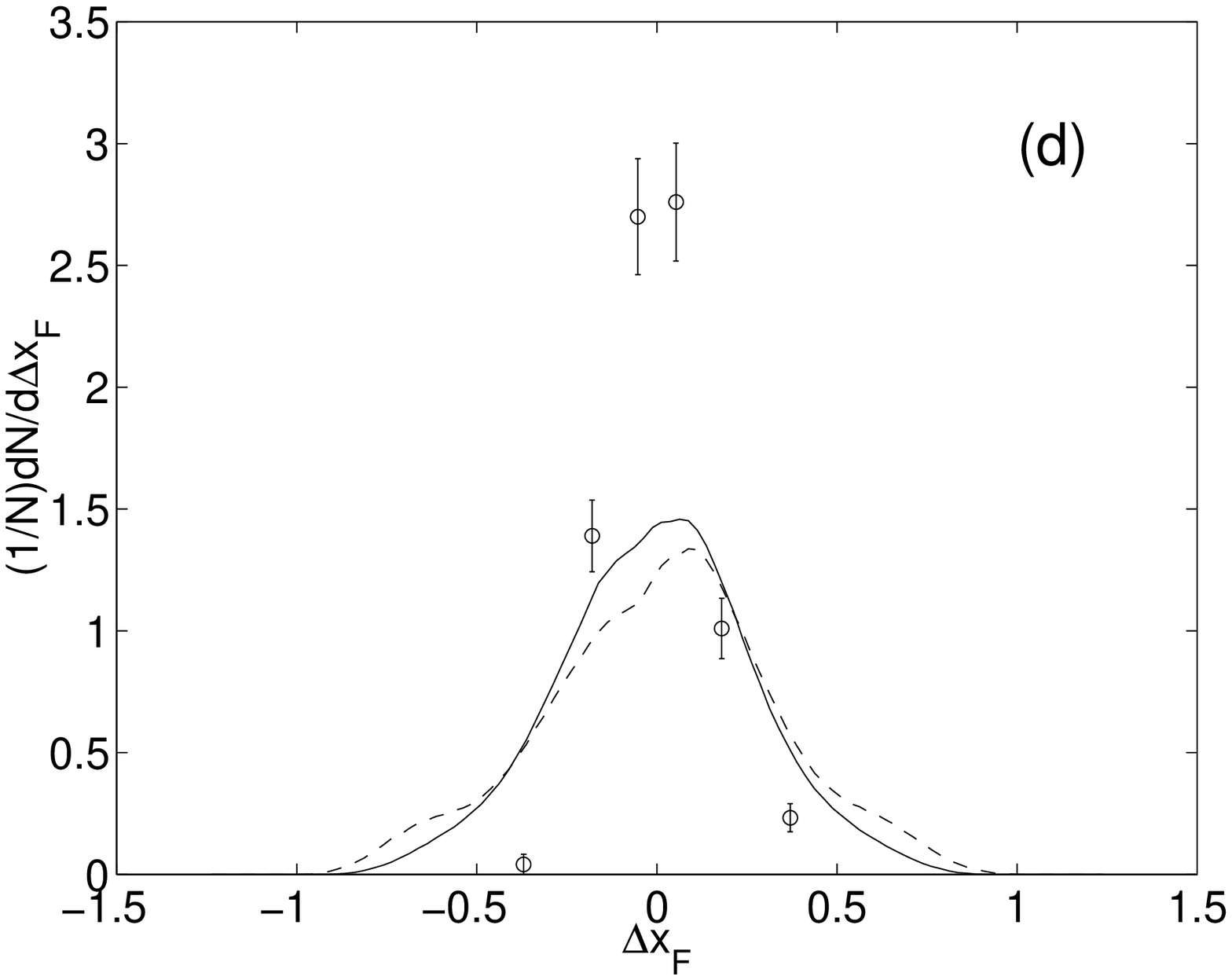, width=48mm}}
\mbox{\epsfig{file=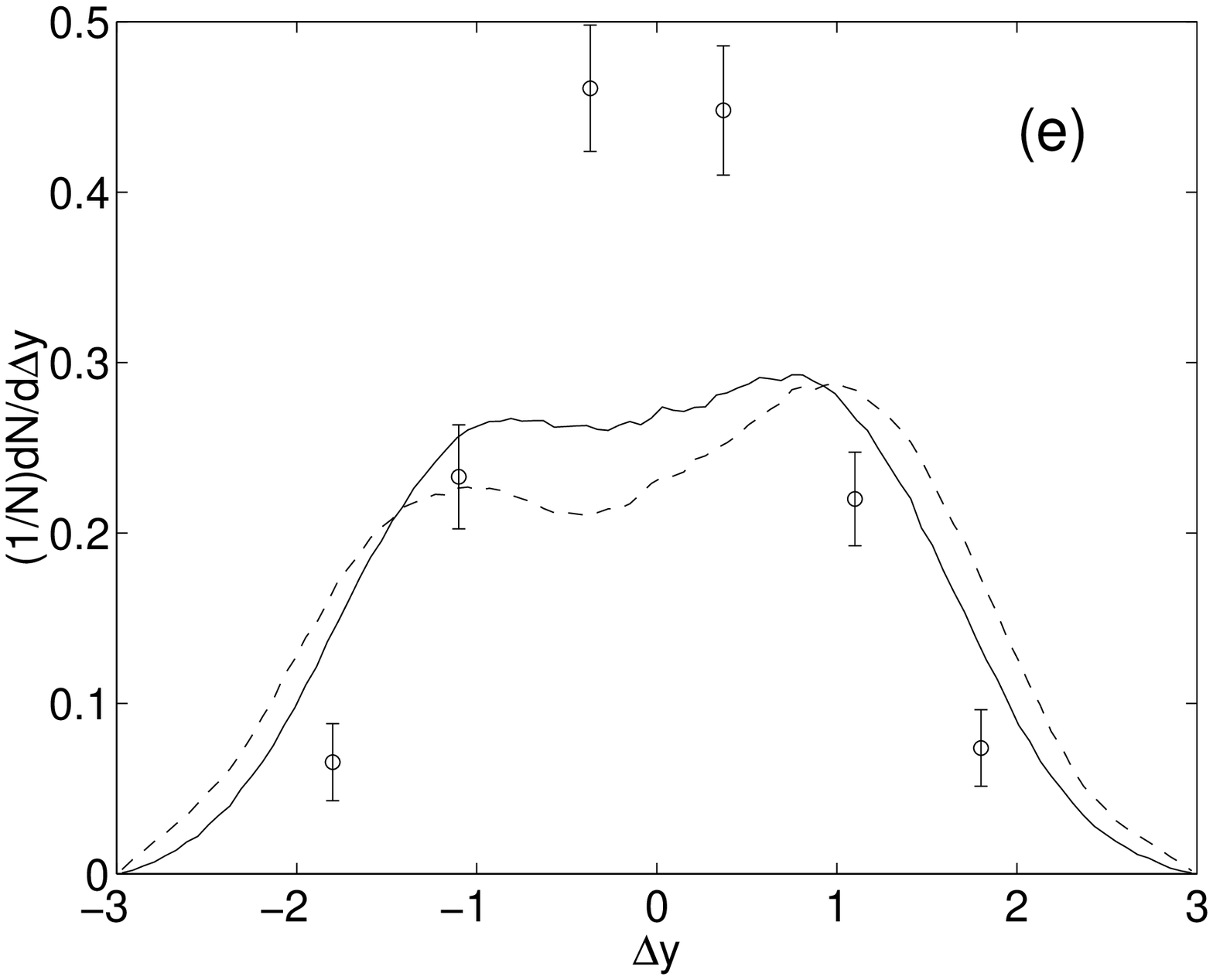, width=48mm}}
\mbox{\epsfig{file=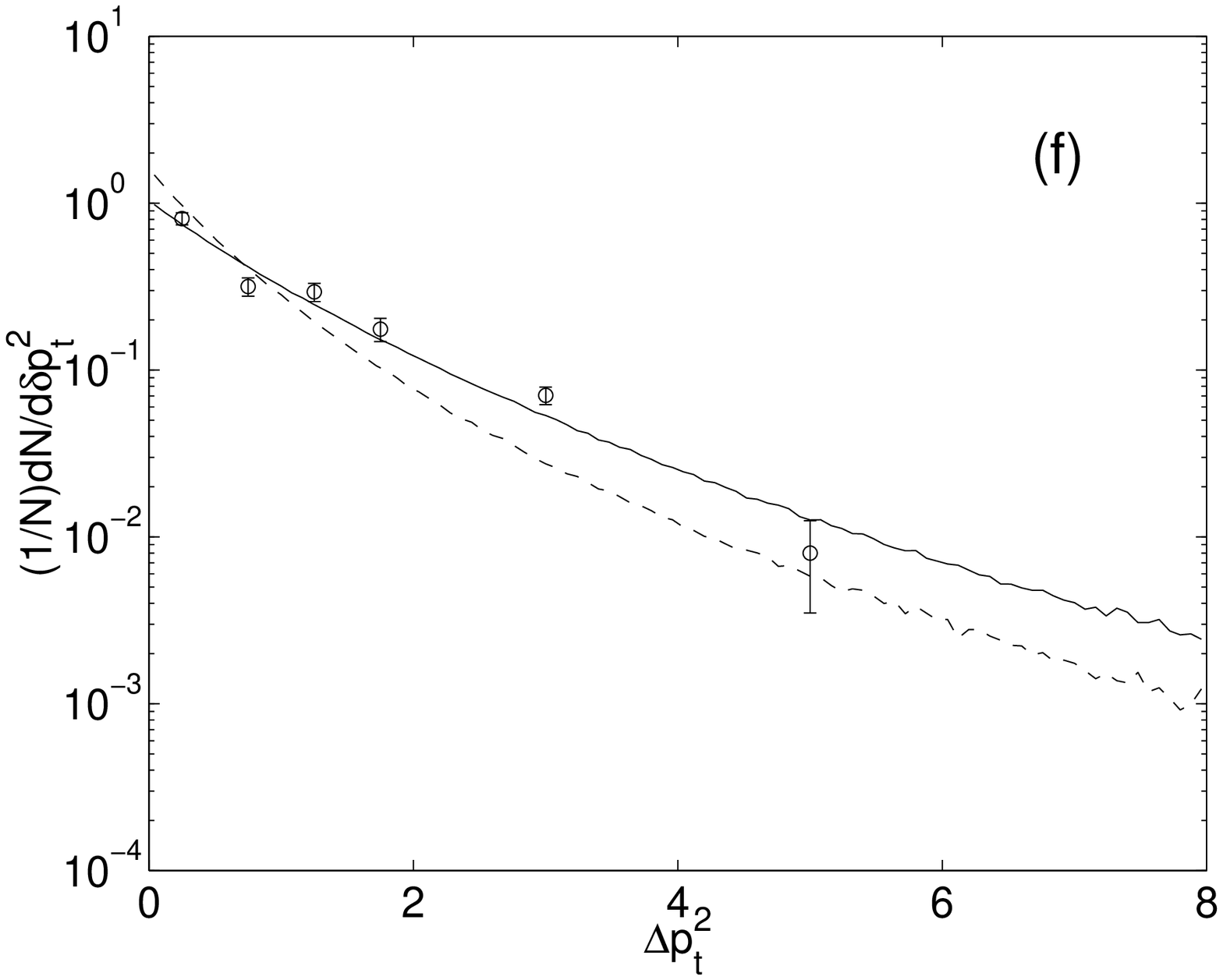, width=48mm}}
\end{center}
\caption[]{$\D\Dbar$ correlations. (a) $\xF$; (b) $y$; (c) $\pt^2$;
(d) $\Delta\xF = x_{\mathrm{F},\D}-x_{\mathrm{F},\Dbar}$;
(e) $\Delta y = y_{\D} - y_{\Dbar}$;
(f) $\Delta\pt^2=|p_{\perp,\D}^2 - p_{\perp,\Dbar}^2|$. Experimental data are from
the E791 experiment \cite{E791corr} compared to the default (dashed) and modified
(full) \Py~predictions.
}
\label{fig.corr}
\end{figure}

Next we compare the model to the new correlation data from E791 \cite{E791corr}.
Several correlations in events where a pair of
$\D\Dbar$ mesons with rapidity in $-0.5<y<2.5$ is fully reconstructed are
studied. The distributions in Fig.~\ref{fig.corr} show that the longitudinal
correlations in the string model are generally different from the data and are better
described by NLO QCD \cite{E791corr}. The transverse correlations on the other hand
are better described by the model, mostly because of the increased primordial $\kt$.

\begin{figure}
\begin{center}
\mbox{\epsfig{file=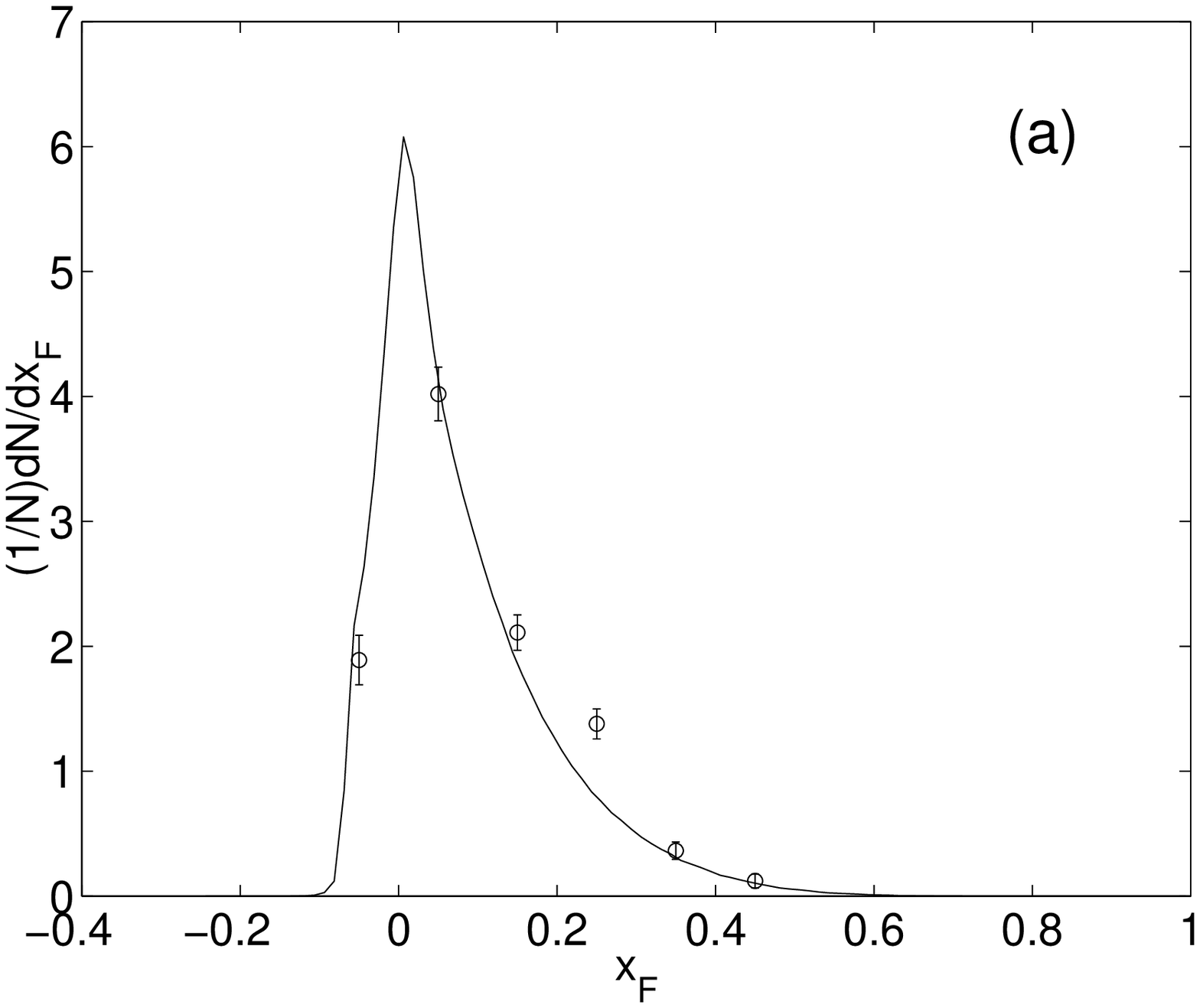, width=48mm}}
\mbox{\epsfig{file=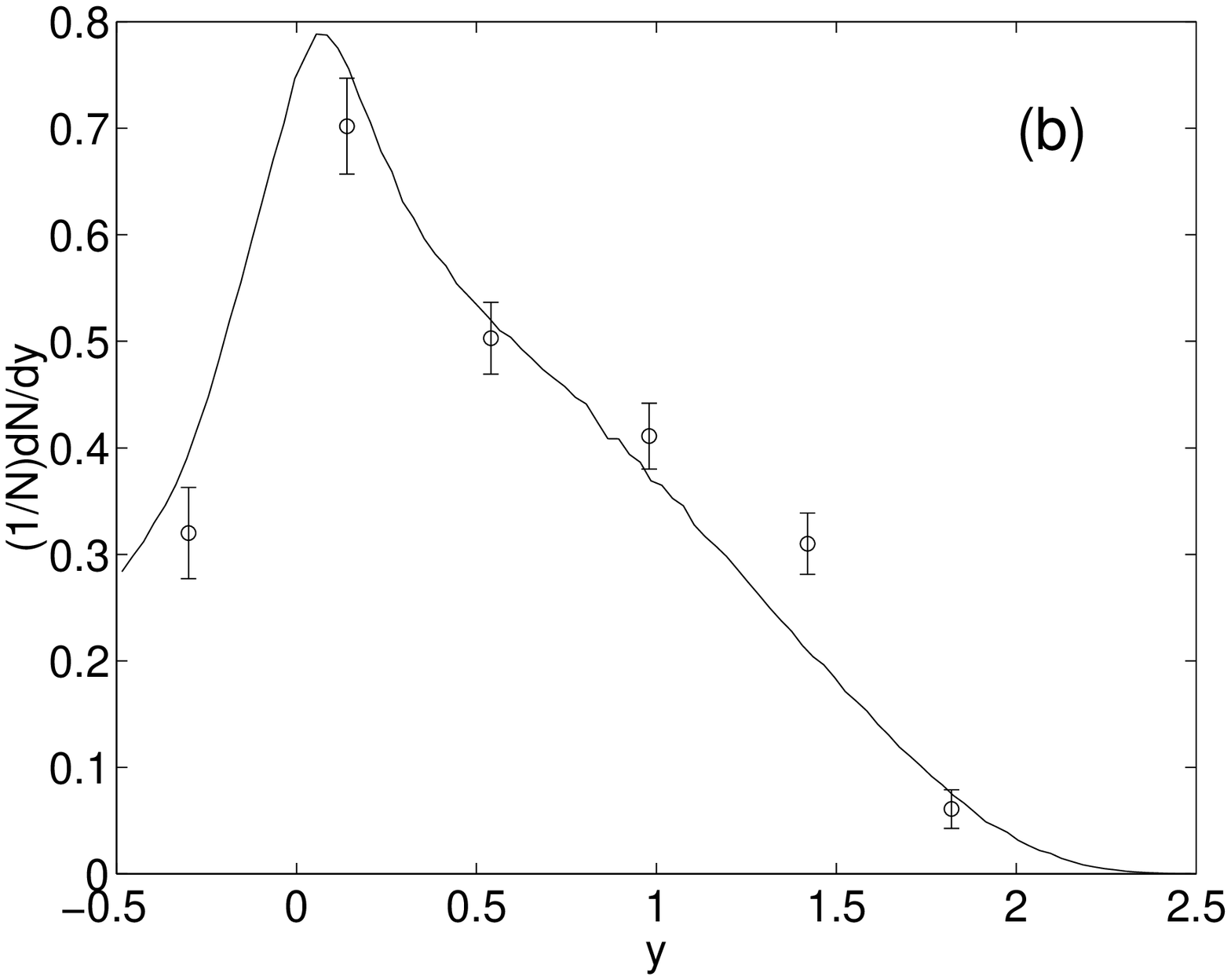, width=48mm}}
\mbox{\epsfig{file=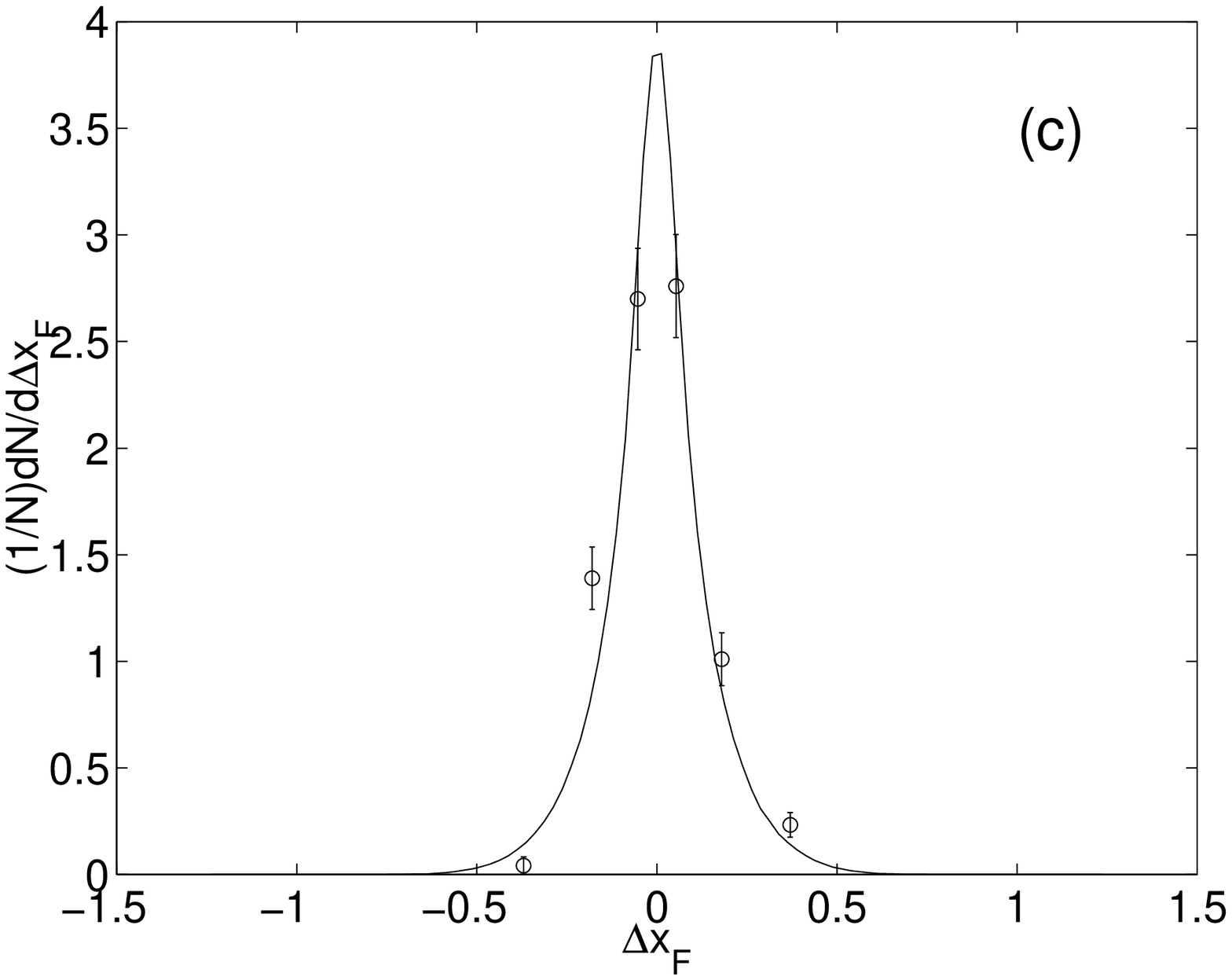, width=48mm}}
\mbox{\epsfig{file=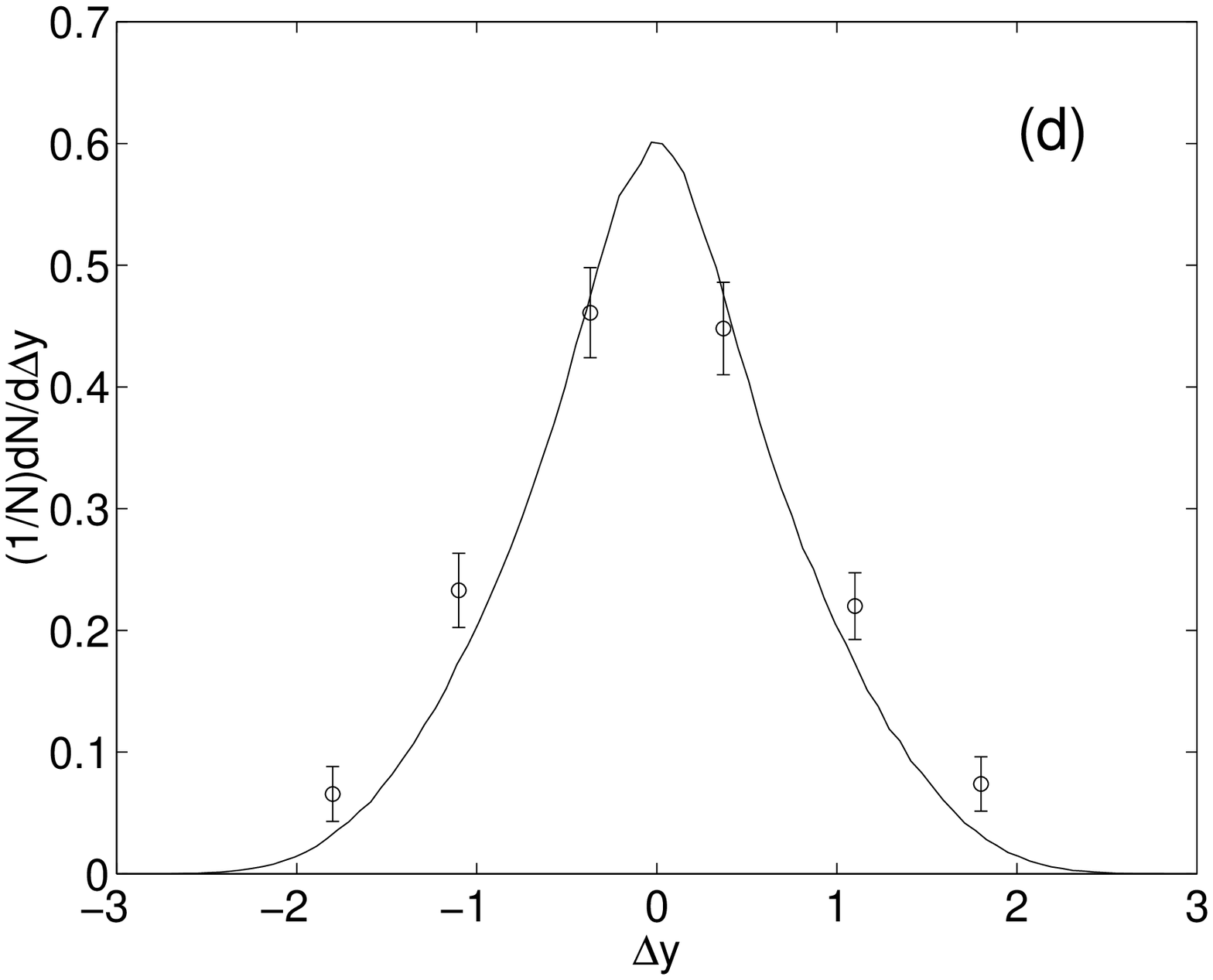, width=48mm}}
\mbox{\epsfig{file=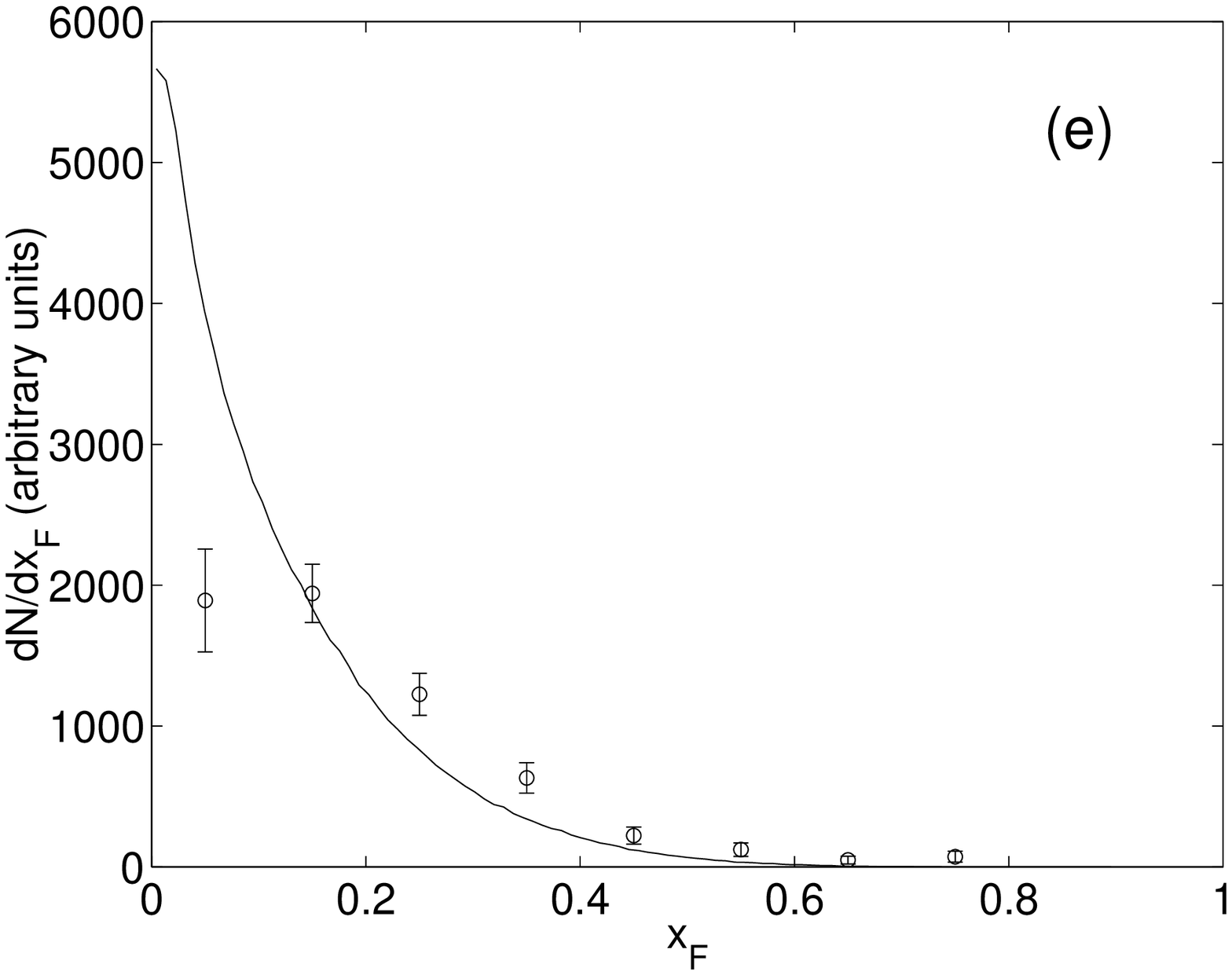, width=48mm}}
\mbox{\epsfig{file=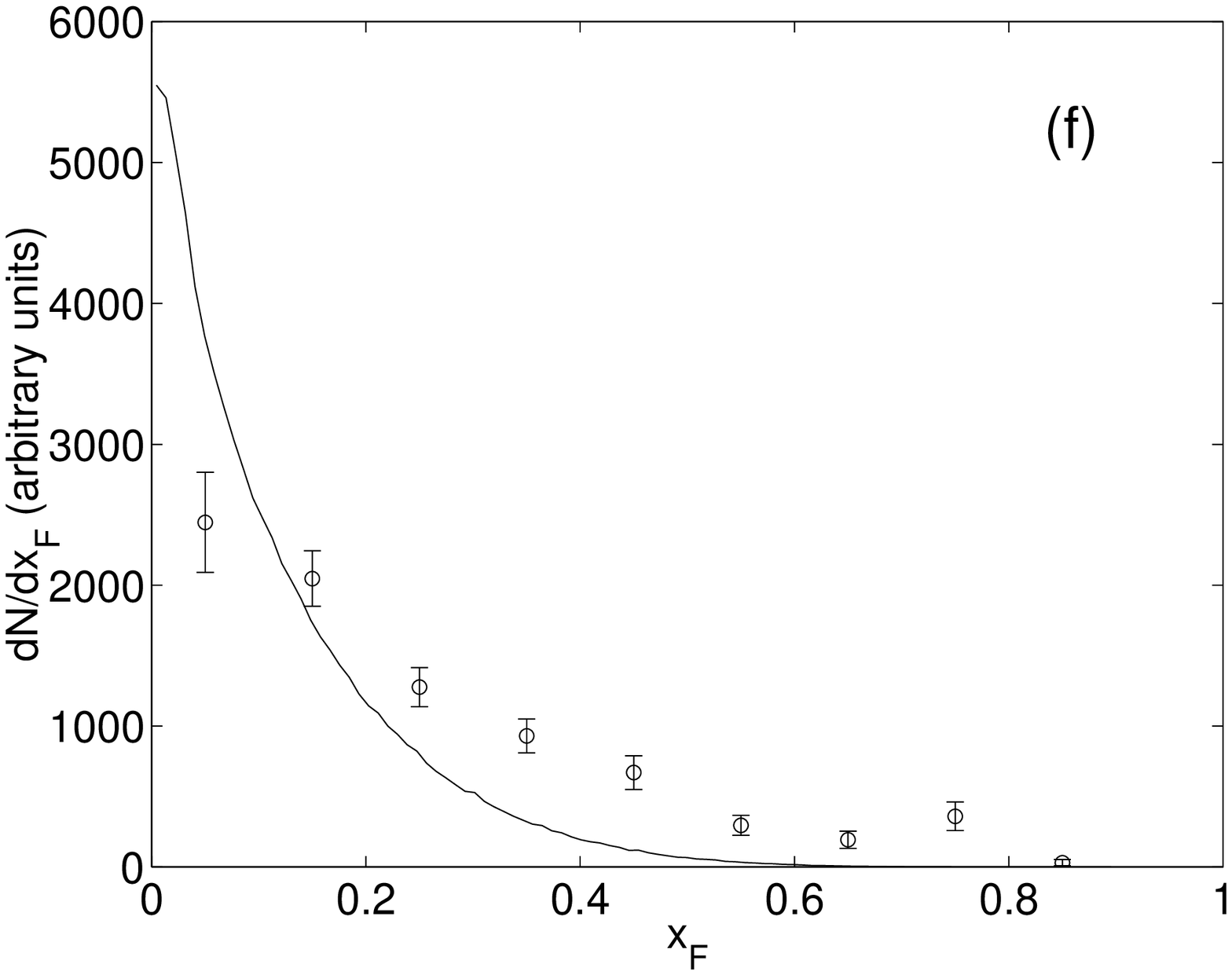, width=48mm}}
\end{center}
\caption[]{D$\Dbar$ correlations compared to independent fragmentation.
Correlations:
(a) $\xF$; (b) $y$;
(c) $\Delta\xF = x_{\mathrm{F},\D}-x_{\mathrm{F},\Dbar}$;
(d) $\Delta y = y_{\D} - y_{\Dbar}$.
Single-charm from WA82:
(e) $\D^+$ and (f) $\D^-$.
}
\label{fig.indep}
\end{figure}

It is clear that the correlation between charm quark pairs should be
modified by hadronization, but the string model seems to produce
D mesons that are too far from each other in momentum (rapidity). To attempt a
description of the data we
use the independent fragmentation approach where a charmed hadron is simply
a scaled-down version of the charm quark. In this way the beam remnants do not
affect the charm quarks at all. We use the fragmentation function
of Peterson et. al. \cite{Peterson} with $\epsilon_{\c}=0.05$. Surprisingly the
longitudinal correlations are quite nicely described by this scheme. Especially the
rapidity distribution is interesting because NLO QCD also fails to describe the strong
peaking at small rapidities seen in the E791 data, see Fig.~\ref{fig.indep}. However,
independent fragmentation of course fails to describe the single-charm spectra from
WA82, falling way below the data and lacking any asymmetry at large $\xF$.
Thus there seems to be a contradiction between the data on single-charm and correlations that
we cannot explain. There are several possibilities that deserve to be investigated.
There could be some cut in the data that we don't understand, or problems
with acceptance corrections at large $\xF$.
Alternatively, some other mechanisms that we have not yet included (sea, intrinsic
charm, parton showers) could give a sizable contribution.

\section*{Summary}

To summarize we have described the string fragmentation
approach to charm production in hadronic collisions. A number of
uncertainties have been identified and studied in detail, in particular
the transition from a continuous string-mass distribution to a
discrete hadron-mass one. The conclusion is that the model can 
describe asymmetries, single-charm spectra and transverse correlations
but not longitudinal correlations. Also, these data do not
fully constrain the choice of model parameters. Further data on
charm production in $\pi^-\p$ collisions may provide further 
information, as may charm production e.g. in $\e\p$ collisions.

\end{document}